\documentclass[journal]{IEEEtran}

\usepackage{cite}
\usepackage{multicol,lipsum}
\usepackage{amsmath,amssymb,amsfonts}
\usepackage{algorithm}
\usepackage{algorithmic}
\usepackage{graphicx}
\usepackage{textcomp}
\usepackage{xcolor}
\usepackage{booktabs}
\usepackage{enumerate}
\usepackage{multirow}
\usepackage{standalone}
\usepackage{tikz}
\usetikzlibrary{positioning,decorations.markings}
\usepackage{float}
\usepackage{caption}
\usepackage{subcaption}
\usepackage{setspace}
\usepackage{verbatim}
\usepackage{tabularx}
\usepackage{makecell}
\usepackage{mathrsfs}
\usepackage{url}
\usepackage{hyperref}

\begin{document}
%
\title{AFDM-ISAC  Design Amalgamating  FMCW Radar and  Carrier Aggregation}

\author{Qu Luo,~\IEEEmembership{Member,~IEEE,} Qihao Peng,  Zihuai Lin,~\IEEEmembership{Senior Member, IEEE},   Maged Elkashlan,~\IEEEmembership{Senior Member, IEEE},  Pei Xiao,~\IEEEmembership{Senior Member, IEEE},  Rahim Tafazolli,~\IEEEmembership{Fellow, IEEE},  and Lajos Hanzo, ~\IEEEmembership{Life Fellow, IEEE}. 
\thanks{The work of Pei Xiao was supported in part by the U.K. Engineering and Physical Sciences Research Council (EPSRC) under Grant EP/P03456X/1 and Grant EP/X013162/1. The work of  Zihuai Lin was supported in part by the Australian Government through the Australian Research Council (ARC) Linkage Project under Grant LP240201020.  The work of Lajos Hanzo  was supported in part by the Platform for Driving Ultimate Connectivity (TITAN) under Grant EP/X04047X/1 and Grant EP/Y037243/1, in part by the Robust and Reliable Quantum Computing (RoaRQ) Project under Grant EP/W032635/1, in part by the PerCom Project under Grant EP/X01228X/1, in part by the EPSRC under Grant EP/Y026721/1, and in part by the India–U.K.  Intelligent Spectrum Innovation ICON under Grant UKRI-1859.  (Corresponding author: Qihao Peng)}
\thanks{Qu Luo, Qihao Peng,   Pei Xiao and Rahim Tafazolli are with 5G \& 6G Innovation Centre, University of Surrey, Guildford, GU2 7XH, UK. (e-mail: {q.u.luo, q.peng, p.xiao, r.tafazolli}@surrey.ac.uk).} 
\thanks{Zihuai Lin is with the School of Electrical and Information
Engineering, University of Sydney, Sydney, NSW 2006, Australia. Email: zihuai.lin@sydney.edu.au; j.yuan@unsw.edu.au.}
\thanks{Maged Elkashlan is with the School of Electronic Engineering and Computer Science, Queen Mary University of London, E1 4NS London, U.K. (e-mail: maged.elkashlan@qmul.ac.uk)} 
\thanks{Lajos Hanzo is with  {the School} of Electronics
and Computer Science, University of Southampton, SO17 1BJ Southampton,
U.K. (e-mail: lh@ecs.soton.ac.uk)}
}

%


\maketitle

\begin{abstract}
Carrier-aggregation (CA)-assisted affine frequency division multiplexing (AFDM) is conceived for integrated sensing and communication (ISAC)  systems. A  frequency-modulated
continuous-wave (FMCW)-inspired dechirping receiver is   developed to exploit the inherent pulse-compression property of AFDM chirp subcarriers for accurate delay–Doppler estimation. To further enhance sensing performance, a pair of CA   schemes  {is} proposed. {The Type-I CA scheme improves sensing accuracy by aggregating echo signals across multiple CA bands prior to radar processing,}  thereby reducing estimation error  {with} low  hardware overhead. The Type-II CA scheme enhances range resolution by effectively enlarging the sensing bandwidth, where chirps from different CA bands are designed to form a continuous set of chirp signals. Based on this principle, variable subcarrier sensing and asynchronous   sensing schemes are developed. For the communication receiver, the generalized input–output relationship of AFDM is derived  under both fractional delay and fractional Doppler (FDFD) channels. To address the challenge of FDFD estimation, a cross-domain sensing-assisted channel estimation (CDS-CE) scheme is proposed  {to} facilitate time-domain  delay–Doppler parameter estimation and    affine-domain channel coefficient recovery. Simulation results demonstrate that the proposed framework achieves significant improvements  {in sensing accuracy and resolution, as well as in channel estimation performance} compared with state-of-the-art methods.
\end{abstract}

\vspace{-0.5em}

%
\IEEEpeerreviewmaketitle

\vspace{-0.6em}
\section{Introduction}

 \IEEEPARstart{I}{ntegrated} sensing and communication (ISAC) has been widely recognized as a key enabling paradigm for next-generation (NG) wireless networks \cite{liu2022integrated,luo2026chirp}. By sharing spectrum, hardware, and signal processing resources, ISAC offers significant advantages in terms of  {spectral} efficiency, hardware cost reduction, and overall system performance enhancement \cite{li2025faster,sui2025multi}. To support ISAC, various radar and communication waveforms have been investigated, among which classic orthogonal frequency division multiplexing (OFDM) has attracted considerable attention due to its widespread deployment in LTE, 5G, and WiFi systems \cite{liu2025cp}. However, emerging high-mobility applications such as smart transportation, the Internet of Vehicles (IoV), and unmanned aerial vehicles (UAVs) impose stringent requirements for waveform robustness against high Doppler shifts and rapid channel variations \cite{rou2024orthogonal}. In such highly dynamic channels, OFDM is   sensitive to carrier frequency offsets and Doppler   spreads, resulting in severe performance degradation \cite{li2022novel,11481148}.

Motivated by the above challenge, several advanced modulation schemes, including orthogonal time–frequency space (OTFS) modulation \cite{raviteja2019effective}, orthogonal delay-Doppler division multiplexing \cite{lin2022orthogonal}, orthogonal chirp division multiplexing (OCDM) \cite{ouyang2016orthogonal},  and affine frequency division multiplexing (AFDM) \cite{bemani2023affine}, have been proposed for robust transmission over    doubly selective channels. Among these candidates, AFDM is a chirp-based multicarrier waveform that modulates information symbols onto a set of orthogonal chirp subcarriers   via the inverse discrete affine Fourier transform (IDAFT). By appropriately tuning the chirp rate according to the channel's delay–Doppler profile, AFDM facilitates a separable and quasi-static channel representation and achieves full diversity in doubly selective channels\cite{luo2024afdm}.  Owing to its strong robustness in high-mobility environments, AFDM has been regarded as a promising waveform candidate for ISAC systems \cite{luo2026chirp,yin2025ofdm}.

\vspace{-1em}
\subsection{State-of-the-art}

AFDM has recently attracted extensive research interest in ISAC waveform design,  ranging from ambiguity function (AF) analysis \cite{ni2025integrated,Yin11173628,ZhangFan11164334,rou2025normalized,bedeer2025ambiguity},   ISAC transceiver design \cite{luo2025novel,luo2025target,ni2022afdm,ZhuAFDMISAC,lu2025daft,ZhangFan11164334,zhu2025isac} and  performance analysis \cite{ma2026mcrb}  to   standardization considerations  \cite{luo2026standardizing}. The AF serves as a fundamental metric for characterizing sensing capability in terms of delay and Doppler resolution.  {Specifically,} Zhang et al.   \cite{zhang2025discrete} investigated the AF behavior of ISAC waveforms carrying random communication symbols.
 It has been shown that while OFDM typically achieves excellent delay resolution and favorable sidelobe characteristics, it suffers from limited Doppler resilience \cite{liu2025cp}. By contrast, AFDM offers a more balanced tradeoff between delay and Doppler resolution, associated with adaptable sidelobe structures depending on parameter selection \cite{ni2025integrated,bedeer2025ambiguity}. Furthermore, the auto-AF of an individual AFDM chirp subcarrier and the cross-AF between distinct AFDM chirp subcarriers were analytically derived in \cite{Yin11173628,ZhangFan11164334}.

Beyond    AF  characteristics, recent studies have further explored AFDM-based ISAC transceiver architectures,  practical sensing parameter estimation strategies and sensing-assisted communication designs \cite{luo2026standardizing}. For example, for radar parameter estimation, a tensor decomposition-based approach was proposed in \cite{luo2025novel}, enabling structured delay–Doppler parameter extraction within the AFDM framework. Moreover, sparse Bayesian learning-assisted methods were developed in \cite{luo2025target} to enhance target detection for high-resolution parameter estimation   in AFDM-ISAC systems.  The
design of sensing methodologies has also been a key area of
investigation. An appealing approach is to employ the 
 full AFDM signal for sensing as in   \cite{ni2022afdm,ZhuAFDMISAC}.  In general, in monostatic sensing where the transmitter and receiver are co-located, the use of full-signal sensing imposes stringent requirements on antenna isolation. Full-duplex operation is   required to suppress the strong leakage from the ISAC transmitter into the receiver chain. To alleviate this issue, deterministic pilot-assisted sensing strategies have been widely investigated \cite{lu2025daft,ZhangFan11164334,zhu2025isac}. For instance, the authors in \cite{lu2025daft} proposed an interference cancellation (IC) scheme in the DAFT domain to mitigate self-interference in pilot-assisted AFDM-ISAC systems.   Additionally,   a single chirp subcarrier-assisted AFDM-ISAC framework combined with maximum-likelihood-based parameter estimation was proposed in \cite{BemaniAFDMisac}.  Furthermore, a dechirping-based AFDM-ISAC receiver was proposed in \cite{Qujstsp}, where the AFDM-ISAC scheme was investigated in a frame-based manner. In \cite{zhu2025isac}, the AFDM parameter-selection criteria were examined, and the mathematical equivalence between AFDM subcarriers and Nyquist-sampled frequency-modulated continuous-wave (FMCW) signals was established.

 {
From a sensing perspective, range and velocity resolutions are fundamentally constrained by the signal bandwidth and coherent processing interval, respectively. Therefore, carrier aggregation (CA)-aided and multiband sensing designs have been investigated to enhance sensing performance. CA has been widely adopted in LTE and 5G NR to improve transmission data rates by allowing a base station (BS) to aggregate multiple fragmented frequency bands \cite{WeiCAISAC,liu2025carrier}. Generally, CA can be classified into three categories, namely intra-band contiguous CA, intra-band non-contiguous CA, and inter-band non-contiguous CA, according to the frequency separation and contiguity of the aggregated component carriers \cite{liu2025carrier}. It has been demonstrated that CA-based ISAC designs  can significantly improve sensing performance in OFDM systems \cite{WeiCAISAC,liu2025carrier}. In \cite{WeiCAISAC}, aggregated pilot structures across different CA bands were designed to enhance sensing performance, whereas in \cite{liu2025carrier}, high- and low-frequency bands were fused by compensating for the initial phase misalignment of the echo signals across the two bands. Moreover, multiband and stepped-frequency radar techniques  have been widely investigated to enlarge the effective sensing bandwidth and thereby improve range resolution \cite{7006701, 10868549, 10838615}. Their basic principle is to transmit or collect sensing signals over multiple frequency bands and jointly process the corresponding echoes to synthesize a wider sensing aperture \cite{10868549, 10838615}. However, these conventional radar designs are primarily sensing-oriented and do not need to account for communication-specific constraints.  To the best of our knowledge,   no results have been reported on CA-assisted ISAC designs in AFDM systems.}

\vspace{-0.6em}
\subsection{Motivation and Contribution}

{Effective exploitation of fragmented spectral resources is essential for simultaneously improving communication throughput and sensing performance in AFDM-ISAC systems \cite{liu2025carrier}. This motivates the investigation of CA-assisted AFDM-ISAC designs that can exploit multiple fragmented frequency bands, while preserving the inherent chirp characteristics of AFDM.}

Against the aforementioned background, the motivations of this work are threefold: 
{1) Although chirp signals have been widely exploited in radar, particularly in FMCW radar, due to their favorable pulse-compression capability, existing AFDM-ISAC studies mainly perform sensing in the affine domain and do not fully exploit the intrinsic chirp pulse-compression  of AFDM subcarriers \cite{luo2025novel,luo2025target,ni2022afdm,ZhuAFDMISAC,lu2025daft,ZhangFan11164334,zhu2025isac}. Therefore, how to leverage these inherent chirp subcarriers for efficient FMCW-like sensing remains largely unexplored; }
 2) Existing AFDM-based ISAC studies primarily focus on sensing design alone, whereas the integration of sensing information to improve communication performance remains insufficiently explored;  {3) Existing CA-assisted OFDM-ISAC schemes mainly perform sensing in the frequency domain and therefore may not be  {well suited to} AFDM-ISAC systems, where sensing is naturally conducted in the affine domain or the time domain (TD) through FMCW-like dechirping processing. }

 The main novelties and contributions of the paper are summarized as follows:
\begin{itemize}
\item We propose a CA-assisted AFDM-ISAC framework in which each CA   band employs a dedicated chirp subcarrier that jointly serves sensing and pilot-assisted communication functions. To fully exploit the \textit{pulse compression} of the  chirp signal, an FMCW-inspired receiver having refined dechirping processing is designed to facilitate accurate delay–Doppler parameter estimation.


 \item To further enhance sensing performance, we propose a pair of carrier-aggregation fusion schemes,  {termed} Type-I CA and Type-II CA.{The Type-I CA scheme improves sensing accuracy by   fusing echo signals } generated from multiple CA bands prior to radar processing.   
 This proposed joint fusion mechanism enables more reliable parameter estimation, while reducing hardware overhead compared to separate sensing on individual bands.  The Type-II CA scheme  {aims to improve} range resolution  by  increasing the effective sensing bandwidth. We show that chirps from different CA bands must form an  {effective continuous} chirp signal set (ECCS) to achieve resolution enhancement. Based on this principle, variable subcarrier sensing (VSS) and asynchronous transmission sensing (ATS) are developed to realize   {ECCS} construction.

 \item Benefiting from the single-pilot-assisted ISAC structure, we further propose a cross-domain sensing-assisted channel estimation (CDS-CE) scheme for fractional delay and fractional Doppler (FDFD) channel estimation. We start by   deriving the generalized input–output relationship of AFDM under FDFD channels. Based on this formulation, the proposed CDS-CE estimates FDFD parameters in the  TD, while the corresponding channel gains are recovered in the DAFT domain.
 
 \item Extensive simulations are conducted to validate the effectiveness of the proposed CA-assisted AFDM-ISAC scheme. The results demonstrate that the proposed Type-I scheme significantly    reduces the mean square error (MSE) of range and velocity estimation. Furthermore, by aggregating multiple CA bands, the Type-II CA scheme is capable of resolving closely spaced targets, thereby enhancing range resolution. In addition, the proposed CDS-CE effectively addresses  FDFD  channel estimation, while achieving superior performance compared to conventional channel estimation methods. 

 \end{itemize}
  \begin{figure*}[htbp]
    \centering
\includegraphics[width=0.9\linewidth]{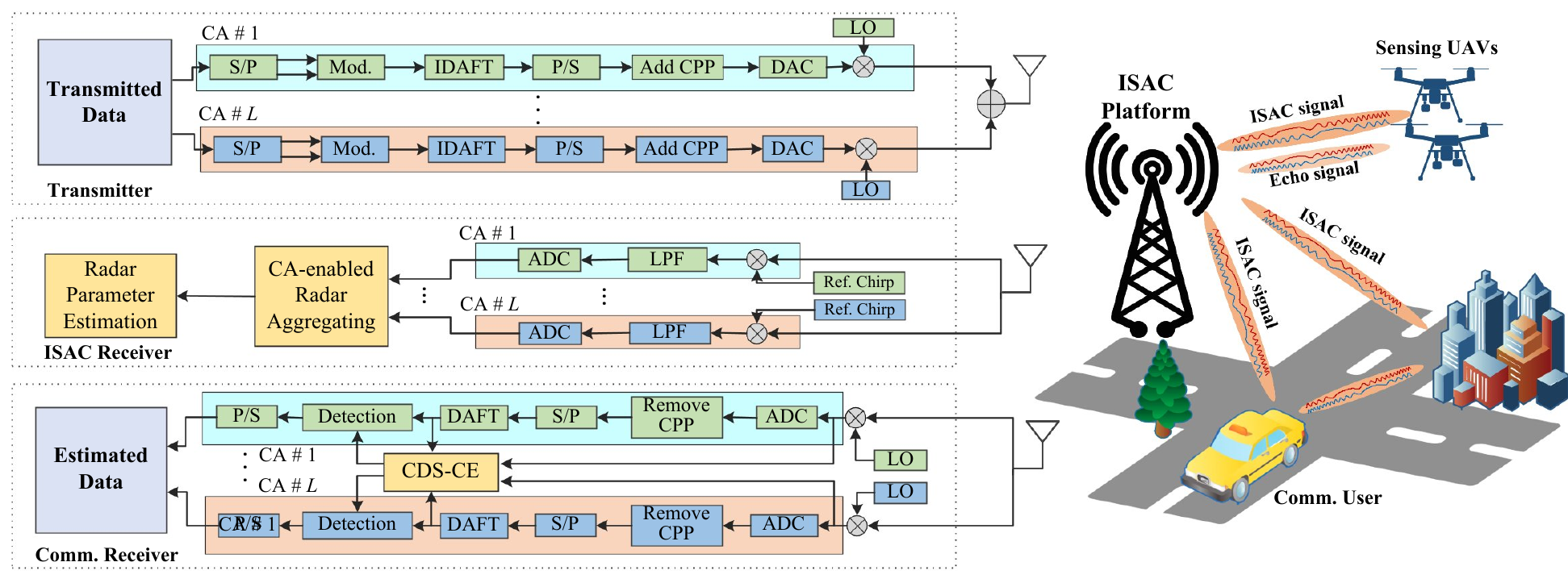}
    \caption{System model of the proposed  {AFDM-ISAC} based on CA (Mod: modulation; DAC: digital-to-analog converter; ADC: analog-to-digital converter; LO: local oscillator; P/S: parallel-to-serial; S/P: serial-to-parallel).}
    \label{fig:system_model}
    \vspace{-1.5em}
\end{figure*}

 The remainder of this paper is organized as follows. Section II introduces the signal model of the CA-assisted AFDM-ISAC system. Section III presents the proposed signal processing framework, including the detailed designs of the Type-I and Type-II CA schemes. Section IV introduces   the proposed CDS-CE scheme. Section V evaluates the sensing and communication performance of the proposed AFDM-ISAC system. Finally, Section VI concludes the paper.
 
\vspace{-1em}
 \subsection{Notation}

  $\mathbb{C}^{k\times n}$ denotes  the $(k\times n)$-dimensional complex  matrix; ${{\mathbf{I}}_{n}}$ denotes an $n \times n $-dimensional  identity matrix;   $\text{diag}(\mathbf{x})$ gives a diagonal matrix with the diagonal vector of $\mathbf{x}$; $(\cdot)^\mathcal T$    and $(\cdot)^\mathcal H$ represent the transpose   and the Hermitian transpose operation, respectively;   $\mathbb{C}$ and $\mathbb Z$ denote the complex and integer spaces, respectively;  $ \left < \cdot \right>_N$ denotes the modulo-$N$ operation,  and   $\mathcal{CN}(0,1) $ is the complex distribution with zero mean and unit variance.

\begin{figure}
    \centering
    \includegraphics[width=0.8\linewidth]{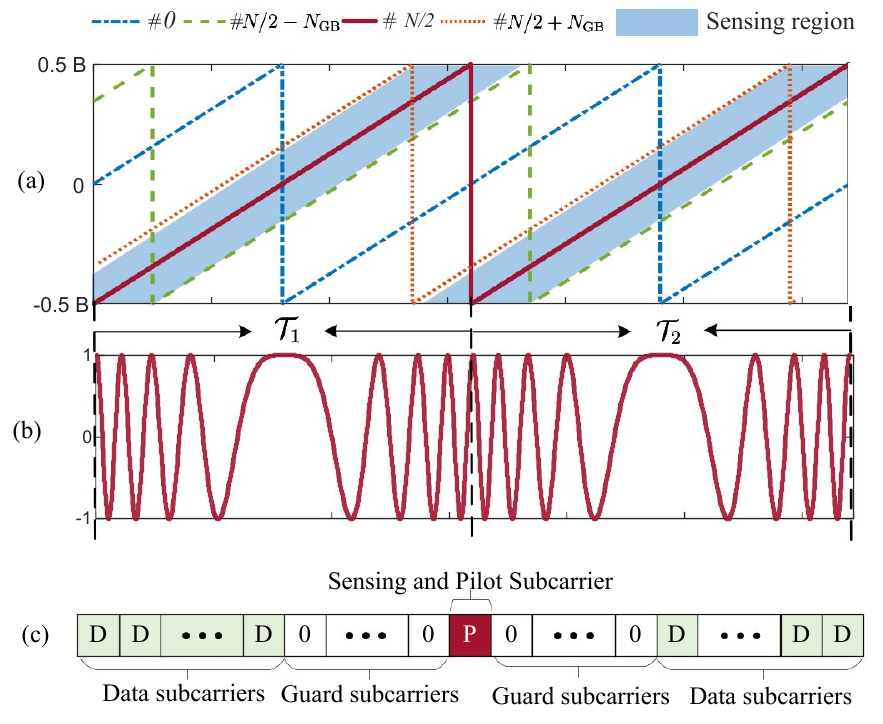}
    \caption{Illustration of (a) the time–frequency representation of chirp subcarriers and the sensing region, (b) the   time-domain SPS, and (c) the sensing and pilot structure, where $c_1 = \frac{2}{2N}$, $N=64$, and $N_{\text{GB}}=10$.}
    \label{PilotStr}
    \vspace{-1.5em}
 \end{figure}






\vspace{-1em}
 \section{System Model of AFDM-ISAC With CA}
 
This section first introduces the  AFDM principles and then presents  {the proposed} AFDM-ISAC model, as shown  in Fig. \ref{fig:system_model}.

\vspace{-1em}
\subsection{AFDM Preliminary}



  The  $n$th AFDM  subcarrier   in the digital-domain is  \cite{bemani2023affine}
\begin{equation}
\Psi_n(m)  ={1\over \sqrt{N}}e^{j2\pi(c_{1}m^{2}+c_{2}n^{2}+{nm\over N})}, 
\label{pjidig}
\end{equation}
where $c_{1}$ and $c_{2}$ are modulation parameters, $0\leq m \leq N-1$ denotes the  TD index,  and $N$ is the number of subcarriers. 
By collecting $N$ normalized samples
of each subcarrier, we
can obtain the so-called discrete Affine Fourier transform  (DAFT) matrix.  Namely, \eqref{pjidig} can be written in   matrix form as   $\mathbf A = \Lambda_{c_2}  \mathbf F \Lambda_{c_1}  $, where $\Lambda_{c} =  \text{diag}\left ( e^{-j2\pi cn^2}, n=0,1, \ldots, N-1 \right)$, and $\mathbf F$ is the  DFT  matrix having the  entries of $[\mathbf F]_{m,n}=e^{-j2\pi mn/N} / \sqrt{N}$. 

Let us denote the subcarrier spacing of AFDM   as  $\Delta f$, and let  $T \triangleq 1/\Delta f$ and $B\triangleq N/T$ be the symbol duration and bandwidth, respectively. Then,  \eqref{pjidig} can be obtained by sampling   the following TD   chirp signal at a sampling rate of $B$: 
  \begin{equation}
  \label{Knt}
 \begin{aligned}
 K_n(t)  =    {{\rm e}^{ j 2 \pi  \left(  {  {\frac{c_1 N^2}{ T^2}}t^{2}  +\frac{  nt}{T}+ c_2n^2  }\right)  }}.
 \end{aligned}
 \end{equation}
For simplicity, we define  the chirp rate as $\alpha \triangleq {\frac{2c_1 N^2}{ T^2}}$.  Observe that \eqref{Knt} occupies a bandwidth of $\alpha T$. According to the sampling theorem, TD sampling introduces spectral wrapping into the digital AFDM signal, thereby limiting the effective bandwidth to the range $[-0.5B, 0.5B]$.  Fig. \ref{PilotStr} (a) shows the time-frequency representation of wrapped AFDM subcarriers.
 
  Upon introducing $\phi_n(t) =  -\frac{\alpha}{ 2}t^{2}  +\frac{  nB}{N}t$,  we have $\phi_n^{'}(t) =  - \alpha t   +\frac{  nB}{N}$. 
The  wrapped AFDM subcarrier in the TD thus can be represented as 
  \begin{equation}
  \label{phint}
 \begin{aligned}
 \Psi_n(t)  =   {{\rm e}^{ j 2 \pi  \left(  { \int_{0}^{t} \bar{{\phi}}_n^{'}(x)dx    + c_2n^2  }\right)  }}, 
 \end{aligned}
 \end{equation}
 where $ \bar{{\phi}}_n^{'}(x) =  \left < \phi_n^{'}(x) + \frac{B}{2}\right>_B-\frac{B}{2}$.



\vspace{-1em}
\subsection{The Proposed AFDM-ISAC  with CA}
 
  In this paper, we assume that $N$ is an even number    and  $c_1 = \frac{Q}{2N}, Q \in \mathbb Z$, i.e., $\alpha = \frac{QN}{T^2}$. From \eqref{phint}, the $\frac{N}{2}$th subcarrier starts from $-0.5 B$ and  {sweeps across} the bandwidth $Q$ times within $T$, while other subcarriers  {sweep across it} $Q+1$ times.  In the proposed AFDM-ISAC, the $\frac{N}{2}$th subcarrier is reserved for   the sensing and pilot subcarrier (SPS).   Based on \eqref{phint}, the $q$th segment of the SPS can   be expressed as    
\begin{equation}
\label{phit}
\begin{aligned}
         \Psi_{\frac{N}{2}}^q(t)  &=   e^{j2\pi(\frac{\alpha}{2} t^2 -(\frac{1}{2}+q)Bt + c_2n^2   )}, t \in \mathcal T_{q},
\end{aligned}
\end{equation}
where $  \mathcal T_{q}  = \Big[\frac{qT}{Q},  \frac{(q+1)T}{Q}\Big ) $, as shown in Fig. \ref{PilotStr} (b).

A double-sided guard having a length of $N_{\text{GB}}$ on each side is considered. The objective of the guard subcarriers is twofold: 1) they essentially correspond to the sensing region and avoid interference caused by communication subcarriers, as illustrated in Fig.~\ref{PilotStr}; and 2) since the $N/2$th subcarrier can also be used as a pilot for channel estimation, the guard regions eliminate interference from neighboring communication subcarriers to the pilot.  Hence, the transmitted signal in the DAFT-domain is given by   
\begin{equation}
  x_n=   \left\{\begin{matrix}
 \sqrt{P_{\text{SPS}}},&   \quad  n= \frac{N}{2},\\
 0, &  0< \vert n-\frac{N}{2} \vert \leq N_{\text{GB}},\\
 \text{Data Symbols,} &  \text{otherwise},\\
\end{matrix}\right. 
\end{equation}
where $P_{\text{SPS}}$ denotes the power allocated to the SPS. We further denote  the set of communication subcarriers as $\mathcal N_{\text{com}}$. 
Accordingly, the $i$th  modulated symbol  in the TD can be expressed as 
\begin{equation}
 s_{i}(t) =   \sum_{n=0}^{N-1}   \text{rect}\big (\frac{t }{T}\big) x_{i,n}  \Psi_n (t), 
 \label{txSt}
\end{equation}
where $x_{i,n}$ denotes the DAFT-domain data at the $i$th symbol of the $n$th subcarrier, and   $\text{rect}(t)$ is the rectangular pulse function.  
A prefix is used to combat dispersion imposed by multipath propagation and  to arrange for the channel  to lie in a periodic domain.   As $c_1 = \frac{Q}{2N}, Q \in \mathbb Z$ is considered, the chirp-periodic prefix (CPP)  of   the AFDM system is essentially a cyclic prefix (CP) \cite{luo2026chirp}.  The TD  transmitted data in the digital domain is expressed as $\mathbf s_i = \mathbf A^{\mathcal H} \mathbf x_i  \in \mathbb C^{N \times 1}$, where $\mathbf x_i  = [ x_{i,0}, x_{i,1}, \ldots,  x_{i,N-1} ]^{\mathcal T}$. Let $N_{\text{CPP}}$ be the length of CPP and define  ${T_{\text{CPP}}=N_{\text{CPP}}T/N}$ and ${T_{\text{sym}}=T+T_{\text{CPP}}}$.  For ease of presentation, the CPP is omitted in \eqref{txSt}, but it will be considered during the  system design.

To improve  SE and support high-data-rate transmission,  CA  is employed. Assuming that $L$ frequency bands are aggregated for transmission, the transmitted signal  is given by
 \begin{equation}
    s_{\text{CA}}(t)  = \sum_{l=0}^{L-1} \sum_{i=0}^{M-1} s_{i}^{l} (t-iT_{\text{sym}}) e^{j2\pi f_c^l t},  
\end{equation}
where  $f_c^l$ denotes the carrier frequency of the $l$th band, $M$ denotes the number of symbols within a frame,  and  $s_{i}^{l}(t)$ denotes the transmitted signal of the $i$th symbol in the $l$th band.{In this paper, we mainly focus on intra-band CA. Since all CA bands are dedicated to the same user and generally experience similar maximum delay and Doppler spreads, the chirp parameters $c_1$ and $c_2$ are assumed to be identical across the $L$ bands.}

\vspace{-1em}
 \subsection{Sensing Channel Model}

The echo signal reflected from the target is a delayed version of the transmitted signal. Similar to existing contributions \cite{luo2025novel,luo2025target,ni2022afdm,ZhuAFDMISAC,lu2025daft,ZhangFan11164334,zhu2025isac }, we assume that the sensing echo is dominated by line-of-sight (LoS) propagation, as illustrated in Fig.~\ref{fig:system_model}.\footnote{ As it will be shown in the proposed CDS-CE scheme, the AFDM-ISAC framework is also capable of processing multiple delay--Doppler components. For sensing, however, multipath components may introduce additional ghost targets. Mitigating such multipath-induced ghost targets generally requires additional target association and multipath separation techniques, which are beyond the scope of this work.} \color{black}    The round-trip delay of the $p$th target is denoted by $\tau_p \triangleq 2R_p/c$, and the corresponding Doppler shift at the $l$th band is given by $f_{\text{d},p}^l \triangleq 2V_pf_c^l/c$, where $c$ denotes the speed of light, while $R_p$ and $V_p$ are the range and velocity of the $p$th target, respectively.
Then, the received echo signal is given by
\begin{equation}
\label{rt}
\begin{aligned}
      r(t)&=\sum_{i=0}^{M-1}\sum_{l=0}^{L-1}r_i^l(t)+ {n(t)}, \\
            r_i^l(t)
      &\triangleq \sum_{p=0}^{P_s-1}a_p s_i^l(t-\tau_p-iT_{\text{sym}})
      e^{j2\pi f_c^l(t-\tau_p)}e^{j2\pi f_{\text d,p}^lt},
\end{aligned}
\end{equation}
where  $P_s$ is the number of sensing targets, $a_p$ denotes the complex gain that models    both channel fading and radar cross-section,        $r_i^{l}(t)$ denotes the received echo signal of the $l$th band  at the $i$th symbol, and $n(t)\sim\mathcal{CN}(0,\sigma_s^2) $ denotes  additive noise.

\vspace{-1em}
\subsection{Communication Channel Model}

For the communication process, a doubly selective channel is considered. Specifically, the channel is assumed to consist of $P_c$ propagation paths, each characterized by a   delay and Doppler shift. Throughout this paper, $\tau$ and the subscript $p$ are reused to denote  the propagation delay and  the $p$th communication path, respectively,    yielding:
\begin{equation}
    h(\tau,\nu) =\sum _{p=0}^{P_c-1} h_{p} \delta (\tau -\tau _{p}) \delta (\nu -\nu _{p}),
\end{equation}
where $h _{p}$, $\tau _{p}$,  and $\nu _{p}$ are the channel gain, delay, and Doppler shift associated with the $p$th path, respectively.  Therefore, the received TD signal before down-mixing can be expressed as 
\begin{equation}
\begin{aligned}
      {y(t)}
      &=\sum_{i=0}^{M-1}\sum_{l=0}^{L-1}\sum_{p=0}^{P_c-1}
      h_p s_i^l(t-\tau_p-iT_{\text{sym}})\\
      &\quad\times e^{j2\pi f_c^l(t-\tau_p)}e^{j2\pi\nu_p^lt}+w(t),
\end{aligned}
\end{equation}
where $w(t)\sim\mathcal{CN}(0,\sigma_c^2)$ denotes   additive noise.

\vspace{-1em}
\section{The Proposed AFDM-ISAC based on CA }


In this section, we first present the proposed  {dechirping-based} sensing scheme, which is compatible with the FMCW receiver. Then, the proposed Type-I and Type-II CA schemes are introduced to improve the sensing MSE and range resolution, respectively. Note that both Type-I and Type-II CA schemes can be implemented simultaneously.

\vspace{-0.8em}
\subsection{Dechirp-aided Sensing}

We start by examining the echo signal corresponding to the desired chirp subcarrier, i.e., the SPS.   Substituting the SPS expression \eqref{phit} into the echo model \eqref{rt} and retaining the $n=N/2$ term,  the received signal corresponding to the transmitted $l$th-band SPS of the $q$th segment can be written as
\begin{equation}
\label{Yinl}
\begin{aligned}
     r_{\frac{N}{2},i}^{l,q}(t_q) & = 
     \sum_{p=0}^{P_s-1} a_p \,
 e^{j2\pi\left(\frac{\alpha}{2}(t_q-\tau_p)^{2}
 -\left(\frac12+q\right)B(t_q-\tau_p  )\right)}
 \\ 
 & \quad \quad \quad  \times  e^{j2\pi f_c^l (t_q-\tau_p)} e^{j2\pi f_{\text{d},p}^{l} t_q}+n(t_q)
 \\
& =  \sum_{p=0}^{P_s-1} a_p   \Psi_{\frac{N}{2}}^q(t_q) e^{j2\pi(\tilde \theta_{p}^{l} + \tilde f_{p}^{l}t_q)} + n(t_q),
\end{aligned}
\end{equation}
where $t_q\triangleq t-iT_{\text{sym}}\in\mathcal T_q $ and 
\begin{equation}
\begin{aligned}
        \tilde {\theta}_{p}^{l} & =\!    \frac{\alpha}{2}\tau_p^2 \!  + \! f_{\text{d},p}^l iT_{\text{sym}} + \left ( \left(\frac{1}{2}+q\right)B \!  - \!  f_c^l \right)\tau_p, \\
 \tilde{f}_{p}^{l} & = \! -\alpha \tau_p + f_{\text{d},p}^l+f_c^l,
\end{aligned}
\end{equation}
are the phase and frequency  {offsets} of the original signal, respectively.  The term $ \sqrt{P_{\text{SPS}}}$  is absorbed into  $a_p $. \color{black}
\begin{figure*}[htbp]
    \begin{equation}
    \label{Mix1}
    \begin{aligned}
         \hat{r}_i^l(t) =    &  \sum_{l'=0}^{L-1} r_i^{l'}(t)s_{\text{ref},i}^{l}(t) 
         = r_i^{l}(t)s_{\text{ref},i}^{l}(t) +   \underbrace{\sum_{l', l'\neq l   }  r_i^{l'}(t)s_{\text{ref},i}^{l'}(t)}_{\text{Inter-BI},   \widetilde {Z}^{l}(t)}  + \tilde{n}(t) \\
    \underset{=}{(\text{i})}   &           \underbrace{  \sum_{p=0 }^{P_s-1}  r_{\frac{N}{2},i}^{l}(t) s_{\text{ref},i}^{l}(t) }_{\text{Desired signal}}   + \underbrace{ \sum_{p=0 }^{P_s-1} \sum_{n \in \mathcal N_{\text{com}}} 
    x_{n,l} a_p \Psi_{n} (t-\tau-iT)  \Psi_{ \frac{N}{2}}^{* } (t-iT)  e^{-2\pi f_c^{l} \tau_p }  e^{2\pi f_p^{l}t}}_{\text{Intra-BI},    {Z}^{l}(t)}    +        \widetilde {Z}^{l}(t)  + \tilde{n}(t) \\
      \underset{=}{(\text{ii})} &  \sum_{p=0}^{P-1} \sum_{q=0}^{Q-1}  \vert a_p  \vert   e^{j2\pi(\theta_{p,q}^{l} + f_{p}^{l}t_q)} +  {Z}^{l}(t) + \widetilde {Z}^{l}(t) + \tilde{n}(t), t_q \in \mathcal T_{q},\\
        \text{where  }  &    {\theta}_{p,q}^{l}  = \phi_p + \frac{\alpha}{2}\tau_p^2 +f_{\text{d},p}^l i T_s + \left ( \left(\frac{1}{2}+q\right)B - f_c^l \right)\tau_p,     \text{ and }   
 {f}_{p}^{l}   = -\alpha \tau_p + f_{\text{d},p}^l.
    \end{aligned}
\end{equation}
 \vspace{-2em}
\end{figure*} 
  On the other hand, the LO of the ISAC  {platform generates} the  reference chirp,  which is the conjugated version of the SPS, i.e., 
\begin{equation}
\begin{aligned}
    &  s_{\text{ref},i}^{l}(t) = \Psi_{\frac{N}{2}}^*\left( t-iT_{\text{sym}}   \right)e^{-j2\pi f_c^{l} t},  l=0,\ldots,L-1.
\end{aligned}
\end{equation}
  {The received echo signal}, i.e.,  $r(t)$ in \eqref{rt},  {is dechirped} with the local signal to remove the pure chirp component.   The detailed expression of the dechirped signal at the $l$th band of the $i$th symbol is given in (\ref{Mix1}), as shown at the top of the next page. In (\ref{Mix1}),  (ii) is obtained by substituting (\ref{Yinl}) into   (i). Furthermore, $ {Z}^{l}(t) $ and $ \widetilde {Z}^{l}(t) $ are the intra-band interference (Intra-BI) and inter-band interference (Inter-BI), respectively, $\phi_p $ is the phase of the channel coefficient, and $\tilde{n}(t)$  denotes the effective noise term. The desired dechirped signal is a sum of complex sinusoids with frequencies of $f_p^l$ and phases $\theta_p^l$. The noise term $\tilde{n}(t)$ is phase-modulated by  $ s_{\text{ref},i}^{l}(t) $, which does not affect the amplitude. 
 
\begin{figure}
    \centering
    \includegraphics[width=0.85\linewidth]{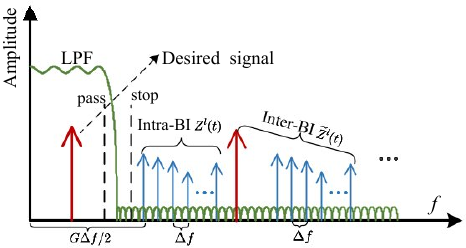}
    \caption{Illustration of the frequency components of the desired signal, inter-BI, and intra-BI, along with the corresponding LPF design.}
    \label{LPFBF}
    \vspace{-1.5em}
 \end{figure}

It should be noted that both  the Inter-BI and Intra-BI have a similar form to that of the desired dechirped signal. 
 Specifically, $ {Z}^{l}(t) $ and   $ \widetilde {Z}^{l}(t) $  cannot be treated as a Gaussian noise term.  Instead, they  {consist of} a sum of complex sinusoids with different frequencies  {from those} of the desired signal.  {Following a derivation similar to step (ii)} of (\ref{Mix1}), one can verify the sinusoidal
 frequencies by dechirping the $n$th subcarrier with $s_{\text{ref},i}^{l}(t)$,   which satisfies
 \begin{equation}
 \label{eqibi1}
  \begin{aligned}
     f_n^l \geq  -\alpha \tau_p + f_{\text{d},p}^l -  (n-\frac{N}{2})\Delta f,&   \quad  n \in \mathcal N_{\text{com}}.
  \end{aligned}
\end{equation}
  Similarly, the dechirped frequency  for   $ \widetilde {Z}^{l}(t) $ satisfies
 \begin{equation}
  \label{eqibi2}
    \widetilde f_n^l \geq  -\alpha \tau_p +f_{\text{d},p}^l -  (n-\frac{N}{2})\Delta f + \underset{\forall l', l' \neq l}{\min } f_c^l-f_c^{l'}. 
\end{equation}
Equations (\ref{eqibi1}) and (\ref{eqibi2}) indicate  that  both $ {Z}^{l}(t) $ and   $ \widetilde {Z}^{l}(t) $   can be filtered out by appropriately designing a low pass filter (LPF) and choosing  a suitable guard band.  Fig.~\ref{LPFBF} illustrates an example of the frequency components of the desired signal, inter-BI, and intra-BI, along with the corresponding LPF design, where $f_{\text{pass}}$ and $f_{\text{stop}}$ are the pass and stop frequencies of the LPF, respectively.

Based on (\ref{eqibi1}), the guard band must be chosen to ensure that $ f_{\frac{N+N_{\text{GB}}}{2}}^l>0$ and $ f_{\frac{N-N_{\text{GB}}}{2}}^l>0$. Namely, 
\begin{equation}
 \label{guard}
      \alpha \tau_{\text{max}} + f_{\text{d},\text{max}}   > N_{\text{GB}}\Delta f, 
\end{equation}
where $\tau_{\text{max}}$ and $f_{\text{d},\text{max}} $ are the maximum delay and Doppler of the sensing target, respectively. {From \eqref{guard}, increasing the guard length can increase the sensing range, but at the cost of reducing the number of subcarriers available for data transmission. Nevertheless, the sensing subcarrier and its surrounding guard region are also reused for sensing-assisted channel estimation, which further mitigates  the resulting communication overhead.}

After passing through the LPF, the dechirped signal is then sampled by the ADC over the measurement interval. Based on \eqref{Mix1}, the sampled data corresponding to the $q$th chirp segment duration of the $l$th band can be expressed as
\begin{equation}
\label{digMix}
\begin{aligned}
        \mathbf {r}_{\text{ADC},i}^{l,q}[m]   
        &\! =  \! \! \sum_{p=0}^{P_s-1}    \vert a_p  \vert   e^{j2\pi\theta_{p,q}^{l}}   e^{j2\pi(f_{p}^{l}mT_{\text{ADC}})} \! + \! \tilde{n}_i(m),  
\end{aligned}
\end{equation}
where $T_{\text{ADC}} = 1/f_{\text{ADC}}$ denotes the ADC sampling interval.  


\vspace{-1em}
 \subsection{The Proposed Type-I CA}
\begin{figure}
    \centering
    \includegraphics[width=1\linewidth]{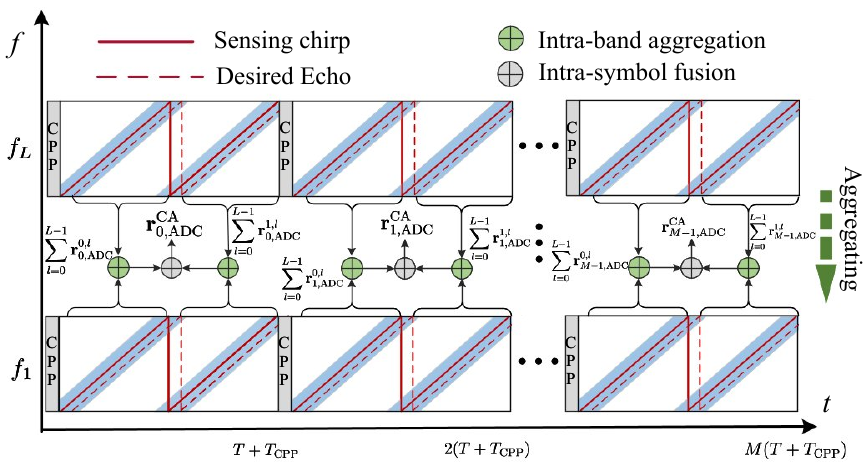}
    \caption{Illustration of the proposed Type-I CA.}
    \label{FrameExamp}
    \vspace{-1.5em}
\end{figure}

This section presents the proposed Type-I CA scheme  {designed to improve} sensing accuracy, as shown in Fig. \ref{FrameExamp}. The objective of the radar receiver in this paper is to estimate the target delay (range) and Doppler shift (velocity) based on \eqref{digMix}. One possible approach is to estimate $f_p^l$ and $f_{\text{d},p}^l$ along the fast-time (i.e., $t_q$) and slow-time (indexed by $i$) dimensions, respectively, for the $L$ bands and $Q$ segments separately. In the proposed Type-I CA scheme, we show that the dechirped signal generated from the $L$ bands, each comprising $Q$ segments, can be judiciously aggregated.

 {As it can be observed from} \eqref{Mix1}, the initial phase term contains additional range and delay information, namely $\frac{\alpha}{2}\tau_p^2$ and $\left( \left(\frac{1}{2}+q\right)B - f_c^l \right)\tau_p$.{For parameter estimation, we distinguish between coherent and noncoherent estimation according to whether the initial-phase information of the dechirped signal is explicitly exploited.}   For the Type-I CA scheme, we propose to merge the dechirped    signals of all CA bands to increase the effective signal power.   
 Note that$-\alpha \tau_p $ is the dominant term of $f_p^l$, i.e., $  \vert f_{\text{d},p}^l \vert  \ll \alpha \tau_p$.   This indicates that the mismatch in $f_c^l$  has less impact on  the estimation of delay and Doppler.  For example, consider two CA bands associated with  $f_c^1 
 = 6$ GHz,  $f_c^2 
 = f_c^1 +\Delta f_{\text{CA}} $ ,  $\Delta f_{\text{CA}}  =100$ MHz, $N =128$, $R = 200 
 \text{ m} $,  $\Delta f =15 \text{ kHz}$, $c_1= \frac{1}{2N}$  and $v = 40  \text { m} / \text s$ then one has 
 \begin{equation}
 \label{deltaF}
\frac{2v\Delta f_{\text{CA}} }{c} = 26.67  \text{Hz}   \ll \frac{2\alpha R}{c} =    3.84 \times 10^{4}  \text{Hz}      . 
 \end{equation} 
In addition, let us define $\Phi^{i,l}_{p,q}(t_q)  \triangleq  \theta_{p,q}^{i,l} + f_{p}^{l}t_q$ as the phase of the desired  term.  The mismatch between  {CA bands} $l$ and $l+1$ in the fast-time dimension is given by 
\begin{equation}
\begin{aligned}
     \Delta^{l \rightarrow l+1}_{\text{d},q} & =    \Phi^{i,l+1}_{p,q}(t_q)  -  \Phi^{i,l}_{p,q}( t_q)  \\
     & =   \frac{2v\Delta f_{\text{CA}} }{c}iT_{\text{sym}} + \frac{2v\Delta f_{\text{CA}} }{c}t_q.   
\end{aligned}
\end{equation}
When estimating the frequency component $f_p^l$ along the $t_q$ dimension, the term $\frac{v \Delta f_{\text{CA}}}{c} iT$ is applied across all frequency bands. {For noncoherent parameter estimation, this mismatch introduces only a constant phase rotation  to all CA bands and therefore does not alter the beat or Doppler frequency. Since range and velocity are estimated from these frequency components, the corresponding 2D-FFT peak locations remain unchanged.}
The same observation holds for the estimation of $\nu_p$ along the slow-time dimension, where the term $\frac{v \Delta f_{\text{CA}}}{c} t_q$ is applied across all frequency bands. This insight therefore motivates us to aggregate the dechirped signals as
\begin{equation}
\label{CA1}
\begin{aligned}
        \mathbf {r}^{q}_{\text{ADC},i}[m] = \sum_{l=0}^{L-1}   \mathbf {r}^{q,l}_{\text{ADC},i}[m],
\end{aligned}
\end{equation}
where the Doppler frequency to be estimated in \eqref{digMix} is rewritten as  $ f_{\text{d},p}=  2v f_c^{\text{CA}} / {c}$ along with   $f_c^{\text{CA}} = \frac{1}{L}\sum_{l=0}^{L-1}f_{c}^l$. Equation \eqref{CA1}  represents intra-band aggregation, as shown in Fig. \ref{FrameExamp}.



To estimate the frequencies $f_p^l$ and $f_{\text{d},p}^l$ along the fast-time   and slow-time  dimensions based on $ \mathbf {R}^{q}_{\text{ADC}} = [ \mathbf {r}^{q, \mathcal T}_{\text{ADC},0},  \mathbf {r}^{q, \mathcal T}_{\text{ADC},1}, \ldots,  \mathbf {r}^{q, \mathcal T}_{\text{ADC},M-1}] $,    existing frequency recovery
algorithms such as MUSIC  or FFT followed
by peak detection can be applied. In this paper, we consider the 2D-FFT associated with constant false-alarm rate (CFAR) detection \cite{kronauge2013fast}.   Finally, instead of applying 2D-FFT  to $ \mathbf {R}^{q}_{\text{ADC}}$ for each segment  (i.e., $q= 0,  \ldots, Q-1$), we propose to merge the $Q$ chirp segments before applying the  2D-FFT detection.      Specifically, the final aggregated  signal, denoted as $   \mathbf {R}_{\text{ADC}}^{\text{CA}}$, can be expressed  as 
\begin{equation}
\label{finalca}
\begin{aligned}
        \mathbf {R}_{\text{ADC}}^{\text{CA}}[i,m] = \sum_{q=0}^{Q-1}  \mathbf {R}^{q}_{ \text{ADC}}[i,m]. 
\end{aligned}
\end{equation}
Equation~\eqref{finalca} remains valid for noncoherent estimation, since the frequencies $f_p^l$ and $f_{\text{d},p}^l$ are identical across all chirp segments within the $i$th symbol. Here, \eqref{finalca} is referred to as intra-symbol fusion, as illustrated in Fig.~\ref{FrameExamp}.{Overall, the proposed Type-I CA-aided sensing procedure is summarized in Algorithm~\ref{alg:TypeI}. }

   \begin{algorithm}[!t]
  
\caption{Proposed Type-I CA-Aided Sensing}
\label{alg:TypeI}
\begin{algorithmic}[1]
\REQUIRE Received echoes $r_i^l(t)$ and reference chirps $s_{\mathrm{ref},i}^{l}(t)$.
\ENSURE Estimated ranges $\{\hat R_p\}_{p=1}^{P_s}$ and velocities $\{\hat v_p\}_{p=1}^{P_s}$.

\FOR{$l=0,\ldots,L-1$}
    \STATE Dechirp $r_i^l(t)$ using $s_{\mathrm{ref},i}^{l}(t)$  {as in} \eqref{Mix1}.
    \STATE Apply LPF and ADC sampling to obtain $\mathbf r_{\mathrm{ADC},i}^{l,q}[m]$.
\ENDFOR

\STATE Aggregate the $L$ CA bands:
$\mathbf r_{\mathrm{ADC},i}^{q}[m]
=\sum_{l=0}^{L-1}\mathbf r_{\mathrm{ADC},i}^{l,q}[m]$.

\STATE Merge the $Q$ chirp segments:
$\mathbf R_{\mathrm{ADC}}^{\mathrm{CA}}[i,m]
=\sum_{q=0}^{Q-1}\mathbf R_{\mathrm{ADC}}^{q}[i,m]$.

\STATE Apply 2D-FFT and CFAR detection to obtain
$\{\hat f_p,\hat f_{\mathrm d,p}\}$.

\STATE Estimate
$\hat\tau_p=(\hat f_{\mathrm d,p}-\hat f_p)/\alpha$,
$\hat R_p=c\hat\tau_p/2$, and
$\hat v_p=c\hat f_{\mathrm d,p}/(2f_c^{\mathrm{CA}})$,
where $f_c^{\mathrm{CA}}=\frac{1}{L}\sum_{l=0}^{L-1}f_c^l$.
\end{algorithmic}
\end{algorithm}

\textit{Remark 1: \eqref{CA1} and  \eqref{finalca} significantly reduce  the observed signal size by fusing the dechirped signal across all CA bands and chirp segments within each symbol. In addition, since the ADC sampling rate is much lower than the AFDM signal bandwidth, the length of   observed data   is substantially smaller than $N$. As a result, the proposed Type-I CA scheme not only achieves a lower estimation complexity owing to the reduced observation data size, but also improves the effective SNR.  }

In the following, we briefly discuss the impact of residual inter-band synchronization impairments, including timing offset (TO), carrier frequency offset (CFO), and phase mismatch, on the Type-I CA aggregation. An inter-band TO occurring before dechirping can be equivalently incorporated into the propagation delay. For the $l$th CA band associated with residual TO $\Delta\tau_l$, the beat frequency becomes
\begin{equation}
\widetilde f_n^l=-\alpha(\tau_p+\Delta\tau_l)+f_{\mathrm d,p}^l,
\end{equation}
where differential TOs across the CA bands cause beat-frequency misalignment. Similarly, a residual CFO, denoted as $\Delta f_l$, mainly introduces a Doppler-frequency mismatch of
\begin{equation}
\widetilde f_{\mathrm d,p}^l=f_{\mathrm d,p}^l+\Delta f_l,
\end{equation}
which mainly degrades Doppler estimation and the effective sensing SNR. By  contrast, a constant inter-band phase mismatch only rotates the dechirped IF signal without changing its beat or Doppler frequency. Hence, it mainly reduces the aggregation gain, rather than shifting the range or velocity peaks of the 2D-FFT. 

It should be noted that, in  the monostatic architecture, the transmitter and sensing receiver are co-located and share a common LO. Therefore, such synchronization impairments are generally less critical than in distributed or bistatic sensing architectures.

\color{black}

\vspace{-1em}
\subsection{The Proposed Type-II CA}


In the proposed Type-II CA radar processing, the objective is to enhance the range resolution by efficiently exploiting the available CA bands. According to the Rayleigh limits  \cite{8955956}, a pair of targets  {is} resolvable, provided that their corresponding frequency components $f_p$ are sufficiently separated. Specifically, for the $p$th and $p'$th targets, the resolvability condition is given by $\delta_f = |f_p - f_{p'}| \ge \frac{1}{T_f}$, where $T_f$ denotes the effective observation duration of the dechirped signal. Hence, it is desirable to increase   ${T_f}$, which is also equivalent to increasing the effective bandwidth of the sensing signal.
Intuitively,  ${T_f}$ can be increased by jointly considering  the  SPS in adjacent CA bands.  In particular, by aggregating the ISAC signals corresponding to the $(i,l)$th and $(i+1,l+1)$st symbols, the effective observation duration $T_f$ can be increased. { {Unlike} existing stepped-frequency radar \cite{10868549, 10838615}, the proposed Type-II CA scheme must account for practical communication constraints, such as guard bands and the CPP, which introduce to the frequency- and time-domain discontinuities in the chirp signals. To address these discontinuities, we first define the  {ECCS} as follows.}

\textit{Definition 1:   Consider a set of  $L$ chirp signals with the same chirp rate $\alpha$, while each signal may have different time and frequency  {durations}.  The  $L$ chirps  form an  {ECCS} if and only if they are partitioned from the same chirp signal.  }

\begin{figure}
    \centering
    \includegraphics[width=1\linewidth]{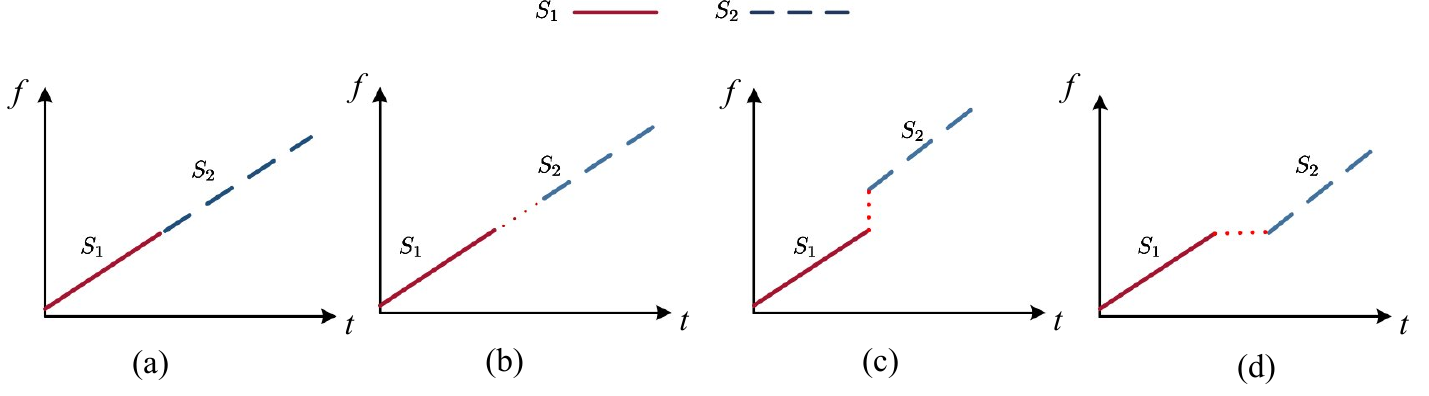}
    \caption{An example of  {ECCS}.}
    \label{ECCSExg}
    \vspace{-1.5em}
\end{figure}


 Fig. \ref{ECCSExg} shows an example of  {ECCS} for $L=2$. Both chirp signals in Figs. \ref{ECCSExg}(a) and  \ref{ECCSExg}(b) can form an  {ECCS}, as they are partitioned from the same   chirp signal.
By contrast, the chirp signals in Figs. \ref{ECCSExg}(c) and \ref{ECCSExg}(d) experience frequency-domain (FD) and TD hops, respectively, which prevent them from constituting an  {ECCS}.
In the following, we analyze the impact of the guard band and the CPP, which give rise to the FD and TD hops, respectively, on the resulting dechirped signal.

For simplicity, assume $Q=1$.  Then the phase difference between   the  symbols at $(i,l)$     and $(i+1,l+1)$     is given by 
\begin{equation}
\begin{aligned}
     \Delta^{l \rightarrow l+1}_{\text{d}}  &  =    \Phi^{l+1}_{p}(i+1,t)\vert_{t = 0}  -  \Phi^{l}_{p}(i,t) \vert_{t = T+T_{\text{CPP}}} \\
     &  =     \underbrace{   - \alpha   T_{\text{CPP}}\tau_p }_{\text{Introduced by  CPP}} + \underbrace{ G^{l\rightarrow l+1 } \tau_p}_{\text{Introduced by  guard band}},    
\end{aligned}
\end{equation}
where $G^{l\rightarrow l+1 }$ denotes the guard band between the $l$th and   $l+1$st bands and $T_{\text{CPP}} = N_{\text{CPP}}T/N$ is the CPP duration.
To ensure phase continuity of the dechirped signal when aggregating the $(i,l)$st and $(i+1,l+1)$th symbols,  {the condition} $ \Delta^{l \rightarrow l+1}_{\text{d}}   \in \mathbb Z$ has to be satisfied.  If the sensing chirps at the $(i,l)$st and $(i+1,l+1)$st symbols constitute an  {ECCS}, it follows that $G^{l \rightarrow l+1} - \alpha T_{\text{CPP}} = 0$, which  yields $\Delta^{l \rightarrow l+1}_{\text{d}} = 0$. However, for non-ECCS, $ \Delta^{l \rightarrow l+1}_{\text{d}}   \in \mathbb Z$  is a   stringent constraint, as $\tau_p, \forall p $  are unknown parameters. 
 
 Based on the above observation,   we propose a pair of schemes, i.e.,  VSS and   ATS.

\textit{\textbf{1) The proposed VSS}}

In the proposed VSS scheme, we relax the constraint of using the $N/2$th subcarrier for sensing.   Define  $n_l$  as the subcarrier selected  from the $l$th band for sensing.  
 To proceed,  the rules of selecting the subcarriers of different bands in the proposed VSS are summarized  as follows:
\begin{itemize}
    \item The   subcarriers selected from different frequency bands for sensing should form an  {ECCS}.  As such, the  {phase discontinuity} of the dechirped signal can be  eliminated.  For $n_l \neq \frac{N}{2}$, each chirp has two segments due to the chirp-wrapping of AFDM. Only the specific segment that  can form an  {ECCS} is selected.  
    \item For each band, it is preferable to select the subcarrier close to the $\frac{N}{2}$th subcarrier, since   the selected chirp segment  {spans a large bandwidth} and thus achieves better range resolution.  More explicitly,  the selected subcarriers should maximize  $\sum_{l} T_{n_l}$, where $T_{n_l}$ denotes the time duration of the selected chirp segment, namely the $n_l$th  {segment}.
    \item    In addition to satisfying  the communication requirements,  $T_{\text{CPP}}$ and the guard band  {also need} to be designed in order to ensure all selected chirp segments form an  {ECCS} and attain high $\sum_{l} T_{n_l}$.
\end{itemize}

\textit{Case-I: Contiguous Band CA:} We first consider $c_1 = \frac{1}{2N}$ as a representative design example, but the proposed procedure can be readily extended to other  
  $c_1$ values. 
In the case of  {contiguous-band} CA, the selected chirp  {subcarriers in adjacent bands}, say the $l$th and $(l+1)$st bands, form an  {ECCS} if the following is satisfied
\begin{equation}
\label{ASSequa}
\begin{aligned}
      G ^{l \rightarrow l+1}+     \left( n_{l+1}-n_{l}\right)\Delta f -  \alpha T_{\text{CPP}} =0, 
\end{aligned}
\end{equation}
where $ 0 \leq n_{l}, n_{l+1} \leq N-1 $. 
Note that both $ G ^{l \rightarrow l+1}$ and $T_{\text{CPP}}$ should be designed to satisfy the      communication requirements and   (\ref{ASSequa}).  Small  values of $ G ^{l \rightarrow l+1}$ and $T_{\text{CPP}}$ indicate better  {spectral} efficiency for communication, while a small value of $\vert n_{l+1}-n_{l} \vert $  leads to a   high effective sensing bandwidth.  Hence, we employ the following metric to guide the design of    $n_l,  G ^{l \rightarrow l+1},$ and $ T_{\text{CPP}}$: 
\begin{equation}
\label{merc}
    \mathcal M_{\text{mer}} = N -   \frac{1}{L-1} \sum_{l=1}^{L-1} \Big \vert  n_{l}- \frac{N}{2} \Big \vert -  N_{\text{CPP}},  L \geq 2. 
\end{equation}
Consequently, the design of the  {ECCS} in the proposed VSS can be formulated as 
\begin{subequations} 
\label{Ometric} 
\small
\begin{align}
 \{n_l,  G ^{l \rightarrow l+1} , T_{\text{CPP}}\}= &   \quad  \arg  \max   \quad       \mathcal M_{\text{mer}}  \tag{\ref{Ometric}}\\ 
     \text{ s.t. } &  \eqref{ASSequa}, \notag  \\    
      & T_{\text{CPP}} \geq T_{\text{CPP}}^{\text{min}},      \label{Ometric b}\\
      &   G ^{l \rightarrow l+1} \geq G^{\text{min}},  \forall l \label{Ometric c} \\
      & n_l \in \{0, 1, \ldots, N-1\},
 \end{align}
\end{subequations}
where  $T_{\text{CPP}}^{\text{min}} =\tau_{\max} $ and  $G^{\text{min}}$ denote the minimum  CPP and guardband   requirements, while $\tau_{\max} $  denotes the  maximum delay spread of communication users.  
 Hence $G^{\text{min}}$  is determined based on the 3GPP  TS 38.101 \cite{3gpp_ts_138101_1_v16_5_0},  {as summarized in} Table \ref{GBR}, for both contiguous and non-contiguous CA.

\begin{figure}
    \centering
    \includegraphics[width=0.9\linewidth]{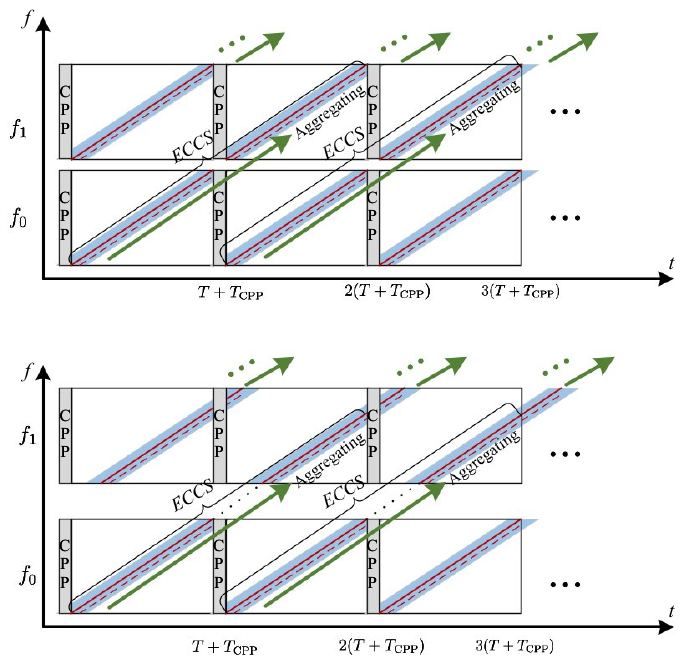}
    \caption{Illustration of the proposed VSS.}
	\label{ifm}
   \vspace{-1.5em}
\end{figure}

It should be noted that if $G^{l \rightarrow l+1}$ and $N_{\text{CPP}}$ are  {chosen to ensure} that  $G^{l \rightarrow l+1} = \Delta f N_{\text{CPP}}$, then (\ref{ASSequa}) can be satisfied with $n_{l+1} - n_{l} = 0$. In this case, both the $l$th and $(l+1)$st bands select the $N/2$th subcarrier for sensing, corresponding to the optimum sensing case. Fig.~\ref{ifm}(a) illustrates an example of this optimum case for $L = 2$. In general, we have $G^{\text{min}} > N_{\text{CPP}}^{\text{min}} \Delta f$, indicating that the optimum sensing condition can always be satisfied by setting $G^{l \rightarrow l+1} = G^{\text{min}}$ and $T_{\text{CPP}} = G^{\text{min}} / \alpha$. However, this configuration introduces a CPP overhead, which may lead to a certain degradation in communication efficiency.

To design the  {ECCS}, we first present the proposed solution to \eqref{Ometric} for   $L = 2$. Specifically, \eqref{merc} can be equivalently expressed as $\mathcal{M}_{\text{mer}} = N - \vert n_{1} - n_{2}\vert - N_{\text{CPP}}$.   The main idea   is to give priority to communication efficiency and to choose $T_{\text{CPP}}$ close to its minimum feasible value.  The detailed procedure is summarized in Algorithm~\ref{ORA}. The    dedicated sensing subcarrier can be determined based on the value of $|n_{1} - n_{2}|$ obtained from the algorithm output.

\textit{Example  2:  We now present  an example  {to characterize} the proposed design.  Consider an ISAC system with system settings of  $f_c = 3.3$ GHz,  $B  =  10$ MHz and $\Delta f = 15$  kHz. Assume that the communication user has a maximum velocity of $200$ km/h, and the maximum delay profile of the EPA channel is considered \cite{luo2024afdm},  {i.e.,} $[0, 30, 150, 310, 370, 710, 1090, 1730, 2510]$ ns.  
Algorithm 1 outputs $N=644$, $N_{\text{CPP}} = N_{\text{CPP}}^{\text{min}} = 25$, $G = 340$ kHz, and $\bar n = 3$. Namely, the $\frac{N}{2}$th and $(\frac{N}{2}+3)$rd subcarriers of the first and second CA bands can be respectively employed for sensing.}

\begin{algorithm} [t]
\small
\caption{ECCS Design for the proposed VSS}
\label{ORA}
 \begin{algorithmic}[1]
\REQUIRE $B, G, N_{\text{CPP}}, G^{\text{min}}, \Delta f$. 
\STATE \textbf{Initialize:}  $N_{\text{ini}}= 2\lfloor \frac{(B-G^{\text{min}})}{2\Delta f}\rfloor$,  $G_{\text{ini}} = B_W - N_{\text{ini}}\Delta f$,  $\bar{n }= n_1-n_2$, $T = 1/\Delta f$.
\FOR{$\bar{n}=-N:N$}
\FOR{$G=G_{\text{ini}}:G_{\text{step}}:G_{\text{max}}$}
\STATE  $N  \leftarrow 2\lfloor \frac{B_W-G}{2\Delta f}\rfloor$,  \quad   $G \leftarrow B_w - N\Delta f$,
  $\alpha \leftarrow \frac{ N\Delta f}{T}$,   
\STATE  \quad $T_{\text{CPP}} \leftarrow  TN_{\text{CPP}}/N$, $G_{\text{res}} \leftarrow G +\bar n  \Delta f - \alpha T_{\text{CPP}} $.
\IF{$\vert G_{\text{res}} \vert <  \Delta f$  }
\STATE $T_{\text{CPP}} \leftarrow T_{\text{CPP}} +    G_{\text{res}}/\alpha$.
\STATE  Preserve $N - \vert \bar n \vert -  N_{\text{CPP}}$  and  the   corresponding    parameter set $\{ N,  T_{\text{CPP}}, G, \bar n  \}$.
\ENDIF
\ENDFOR
\ENDFOR \\
\ENSURE{ Output $\{ N,  T_{\text{CPP}}, G, \bar n  \}$ that achieves the maximum value of $N - \vert \bar n \vert -  N_{\text{CPP}}$.}  
 \end{algorithmic}
\end{algorithm}

  \begin{table}[t]
  \footnotesize
 \centering
 \caption{  \small
{Minimum guardband (kHz) specified in 5G NR.}}
     \begin{tabular}{c|c|c|c|c|c}
    \hline
    \hline
        \makecell{ $\Delta f$\\
        \&  $B$}    &  $5$ MHz  & $10$ MHz &  $15$ MHz & $20$ MHz  &  $25$ MHz \\
    \hline
    \hline
    15  & 242.5  & 312.5 & 382.5 &452.5 & 522.5\\
    \cline{1-6}
 30   & 505  & 665 &645 &805 & 785\\
     \cline{1-6}
 60   & N/A & 1010& 990 &1330 & 1310\\
   \cline{1-6}
    \hline
    \hline
     \end{tabular}
     \label{GBR}
 \end{table}

For $L > 2$, the selection procedure differs for odd and even values of $L$. For an odd $L$, the $N/2$th chirp subcarrier of the $(L/2 + 1)$st band is first dedicated to sensing. The chirp subcarriers of the adjacent bands, along with their corresponding parameters, are then determined using Algorithm~\ref{ORA}. Finally, the chirp subcarriers of the remaining bands are obtained by recursively applying Algorithm~\ref{ORA}.
For an even $L$, the chirp subcarriers corresponding to the $(N/2 + 1)$st and $(N/2 - 1)$st bands are jointly determined using Algorithm~\ref{ORA}. Subsequently, the  {ECCS} is constructed by recursively determining the chirp subcarriers of the adjacent bands. 

\begin{figure}[t]
    \centering
\includegraphics[width=0.88\linewidth]{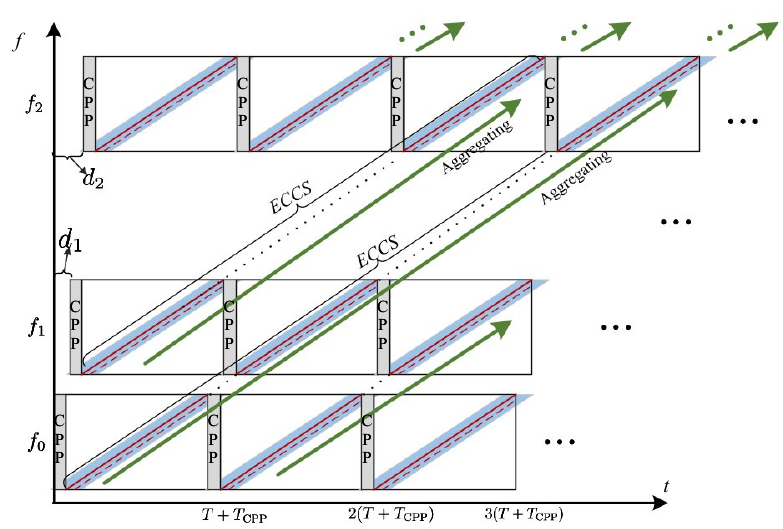}
    \caption{ Illustration of  the proposed ATS.}
    \label{BlankingExam}
    \vspace{-1.5em}
 \end{figure}

\textit{Case-II: Non-Contiguous  Band:}  For a non-contiguous band,  assume that the  $n_l$th chirp of the $i_l$th symbol, and the $n_{l+1}$th chirp   of the $i_{l+1}$th  symbol are selected to  form an  {ECCS}. Then, similar to (\ref{ASSequa}), the following needs to be  satisfied
\begin{equation}
\label{ASSequa2}
\begin{aligned}
      G ^{l \rightarrow l+1} \!+  \!\alpha(i_{l+1}-i_{l})(T+ T_{\text{CPP}}) \!+ \! \left( n_{l+1}-n_{l}\right)\Delta f \!-\! \alpha T_{\text{CPP}} =0. 
\end{aligned}
\end{equation}
In order to satisfy   (\ref{ASSequa2}), we first set $i_{l+1}-i_{l}=\mod (f_{l+1}-(f_l+B), B)$. Then, the rest of the parameters can be readily obtained by using Algorithm \ref{ORA}.

\textit{\textbf{2) The proposed ATS}}

By  contrast to the VSS scheme, the proposed ATS scheme utilizes the $N/2$th subcarrier for sensing in all CA bands.    To form an  {ECCS}, the BS transmits signals from  different CA bands with certain delays. Specifically, the BS transmits the signal of the $l$th band with a delay of $d_l$, where $0 \leq d_l <T$,  at the  beginning of each frame. Let the   $N/2$th subcarrier of the $i$th symbol at the first CA band be the $i$th  {ECCS} group, and assume $d_0=0$.  Based on the  {ECCS} principle, the following should hold
\begin{equation}
\label{ATSs}
\begin{aligned}
 \alpha  (d_{l} -d_{l-1})    = G ^{l-1 \rightarrow l},  \quad     l=1, 2,\ldots, L-1,  
\end{aligned}
\end{equation}
which can be sequentially  {solved} for $ l=1, 2,\ldots, L-1$. 
Fig. \ref{BlankingExam} shows an example of the proposed ATS for $L=3$. The  {first and second bands} are two adjacent bands, but the third band is located farther away from the second band in the FD.   As  seen from the figure, each band has a delay of $d_l$. As such, the $N/2$th subcarrier in  the $i$th symbol of the first band, the $(i+1)$st symbol of the second band, and the $(i+3)$rd symbol of the third band form the $i$th  {ECCS}. 
 Recall   the dechirped signal in \eqref{digMix}, i.e., $\mathbf{r}_{\text{ADC}}^{l,q}$.   The proposed ATS aggregates  the dechirped signal   as
\begin{equation}
   \mathbf {r}_{i,\text{ADC}}^{q,\text{ATS}} =  \left [   \left [\mathbf {r}^{q,0}_{i_0, \text{ADC}}\right ]^{\mathcal T},    \ldots,   \left [\mathbf {r}^{q,L-1}_{ i_{L-1}, \text{ADC}} \right ]^{\mathcal T}    \right ]^{\mathcal T},
\end{equation}
where $\mathbf {r}_{i,\text{ADC}}^{q,\text{ATS}}$ is the aggregated signal of the $i$th  {ECCS}.  By collecting $M$ aggregated  {ECCSs}, the resultant signal  is used for sensing parameter estimation.

\begin{figure*}[tbp]
     \centering
     \includegraphics[width=0.9 \linewidth]{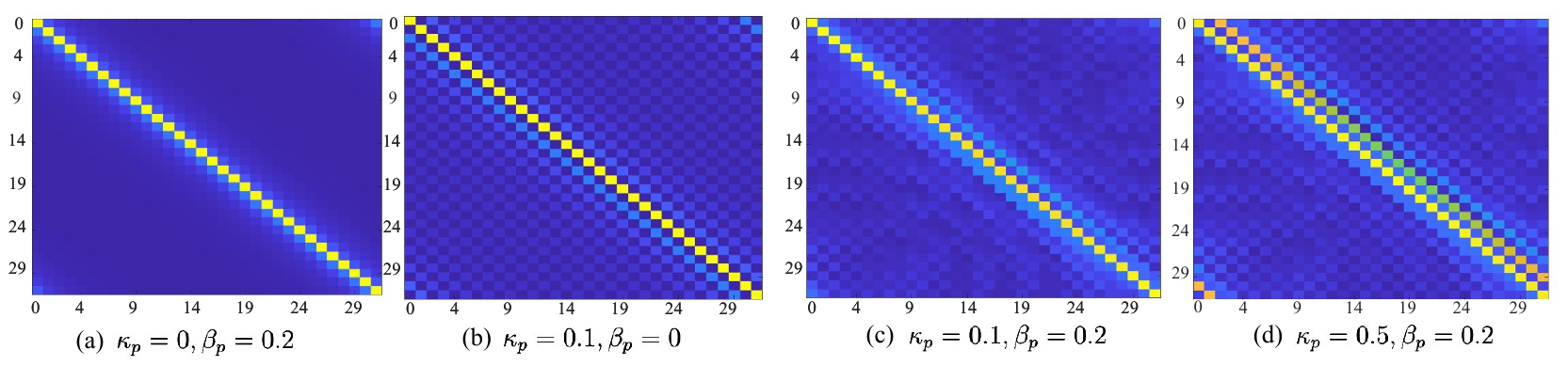}
     \caption{Illustration of  {the} effective channel  with (a) $\kappa_p^{\text{F}} = 0, \beta =0.2$, (b) $\kappa_p^{\text{F}} = 0.1, \beta =0$,  (c) $\kappa_p^{\text{F}} = 0.1, \beta =0.2$, (d) $\kappa_p^{\text{F}} = 0.5, \beta =0.2$, where $N = 32$ and $Q=2$.}
     \label{ChR}
    \vspace{-1.5em}
 \end{figure*}

\begin{figure}
     \centering     \includegraphics[width=0.85\linewidth]{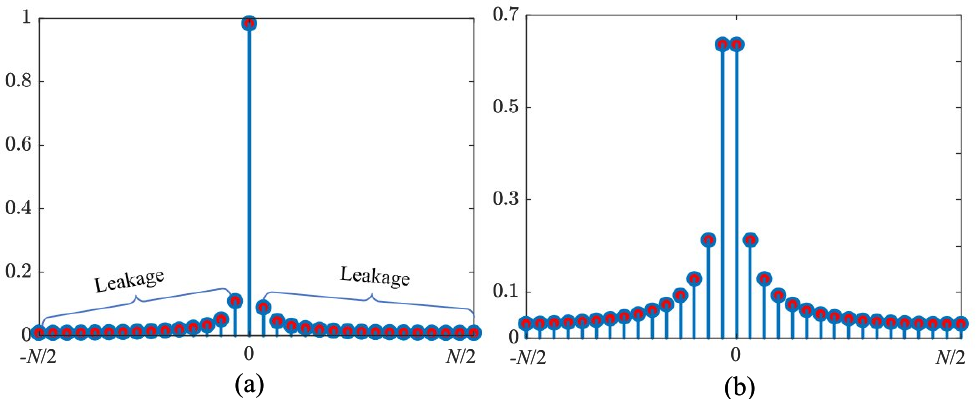}
     \caption{Illustration of  {fractional} delay  with (a) $\kappa_p^{\text{F}} = 0.1$ and (b) $\kappa_p^{\text{F}} = 0.5$.}
     \label{FracDel}
     \vspace{-1.5em}
  \end{figure}

\vspace{-1em}
\section{Sensing-assisted Communication Receiver Design}
This section introduces the proposed sensing-assisted receiver design, i.e., CDS-CE. We first derive the input–output relationship of AFDM for transmission over FDFD channels in Section~\ref{AFDMFDFD}. Then, Section~\ref{PSCE} presents the proposed CDS-CE scheme to address the challenges associated with FDFD estimation.

\vspace{-1em}
\subsection{AFDM over FDFD Channels}
\label{AFDMFDFD}
We define   $\tau_p \triangleq \frac{\kappa_p^{\text{I}}+\kappa_p^{\text{F}}}{N\Delta f} $ and $\nu_p \triangleq \frac{\beta_p^{\text{I}}+\beta_p^{\text{F}}}{T} $, where $\kappa_p^{\text{I}}$, $\beta_p^{\text{I}}$  are  integers and  $\kappa_p^{\text{F}}$, $\beta_p^{\text{F}}$ are real values satisfying  $-\frac{1}{2} \leq \beta_p^{\text{F}}, \kappa_p^{\text{F}} < \frac{1}{2}$. Namely,  $\kappa_p^{\text{F}}$ and  $\beta_p^{\text{F}}$  are the fractional parts of the delay $\tau_p $ and Doppler $\nu_p$, respectively.  Existing AFDM systems generally consider delays on the integer grid. Specifically, the  signal in the DAFT-domain received with integer delay and fractional Doppler   can be expressed as follows:
\begin{equation}
\label{heff}
\begin{aligned}
        \mathbf {y} &= \mathbf {H}_{\text{eff}} \mathbf {x} + \tilde {\mathbf {n}} \triangleq \sum _{p = 0}^{P_c-1}h_{p}   \underbrace{\mathbf A \boldsymbol{\Gamma}_{\mathrm {CPP}_{p}} {\boldsymbol{\Delta }}^{\beta_{p}} {\boldsymbol{\Pi }}^{\kappa_p^{\text{I}} }\mathbf A^{\mathcal H} }_{\mathbf {H}_{p}} \mathbf {x} + \tilde {\mathbf {w}},
\end{aligned}
\end{equation}
where $ \mathbf {H}_{p} $ is the  effective channel matrix of the $p$th path. Furthermore,  $\beta_p = \beta_p^{\text{I}}+\beta_p^{\text{F}}$, while     $\boldsymbol {\Delta}^{\beta_p} = \text{diag}(e^{-j \frac{2\pi\beta_p}{N} n  }, n=0,1,\ldots,N-1)$ models the Doppler effect,  
   $\mathbf  \Pi $
 denotes the forward cyclic-shift matrix,   and 
 \begin{equation}
    \mathbf{\Gamma}_{\text{CPP}_p}    = \text{diag} \left( \left\{\begin{matrix}
  e^{-j2\pi c_1 (N^2-2N(\kappa_p^{\text{I}}-n))}, &   n< \kappa_p^{\text{I}}, \\ 
  1, & n \geq \kappa_p^{\text{I}},
 \end{matrix}\right. \right) 
 \end{equation}
     models the  {CPP}.

 \begin{figure*}[t]
  \begin{equation}
  \label{finalH}
\begin{aligned}
    \big [  \mathbf H_p \big ]_{u,v} &  = \left [ 
    \sum_{k=0}^{N-1} D_{N} \Big ( \frac{k-\kappa_p^{\text{F}} }{N} \Big)  \mathbf A  {\boldsymbol{\Gamma }}_{\mathrm {CPP}_{p}}  {\boldsymbol{\Delta }}^{\beta_p}  \boldsymbol \Pi^{k+\kappa_p^{\text{I}}} \mathbf A^{\mathcal H}\right ]_{u,v}  
    = \sum_{k=0}^{N-1} D_{N} \Big ( \frac{k-\kappa_p^{\text{F}} }{N} \Big) \sum_{n=0}^{N-1}  \big [ \mathbf A \big ]_{u,n} e^{-j2\pi \frac{\beta_p  n}{N}  }  \big [ \mathbf A \big ]_{v,n-k-\kappa_p^{\text{I}}  }^{*} \\
    &=  e^{j2\pi c_2(v^2-u^2) }  \sum_{k=0}^{N-1}  \underbrace{  D_{N} \Big ( \frac{k-\kappa_p^{\text{F}} }{N} \Big) e^{ j \frac{2 \pi}{N}  \left (Nc_1 (k+\kappa_p^{\text{I}})^2-  v(k+\kappa_p^{\text{I}})  \right )  }}_{ k\text{th tap  fractional delay},  \mathcal F_{p}^k(v)  }  \underbrace{ D_{N}\Big (    \frac{v-u-\beta_p - 2Nc_1(k+\kappa_p^{\text{I}})}{N}  \Big)}_{\text{Doppler spread}, \mathcal D_{p}^k(u,v)}   
\end{aligned}
\end{equation}
  \par\noindent\rule{\linewidth}{0.4pt}
  \vspace{-1em}
 \end{figure*}

It is worth noting that the diagonal Doppler matrix $\boldsymbol {\Delta}^{\beta_p}$, which consists of exponential terms, can inherently represent both integer and fractional Doppler shifts. By contrast, the circulant delay matrix in $ \mathbf {H}_{p} $  is incapable of directly modeling fractional delays. To address this limitation, the circulant delay matrix can be substituted with a diagonal delay matrix, where the circular shifts corresponding to target-induced delays are mapped into FD phase shifts through the DFT matrix. Applying the inverse DFT matrix then restores the signal to the TD.   Explicitly, we have:
\begin{equation}
\label{Hpt}
    \mathbf {H}_{p} =  \mathbf A  {\boldsymbol{\Gamma }}_{\mathrm {CPP}_{p}}  {\boldsymbol{\Delta }}^{\beta_p}  \mathbf F^{\mathcal H} \boldsymbol{\Gamma}^{\kappa_p}  \mathbf F\mathbf A^{\mathcal H},    
\end{equation}
where $\boldsymbol{\Gamma}^{\kappa_p} = \text{diag}\big(e^{-j\frac{2\pi\kappa_p }{N}n}, n = 0,1,\ldots, N-1 \big )$ and  $\kappa_p = \kappa_p^{\text{I}}+\kappa_p^{\text{F}}$. The term $\mathbf F^{\mathcal H} \boldsymbol{\Gamma}^{\kappa_p}  \mathbf F$ reduces to the conventional circulant delay matrix for any integer delay.   It can be shown that 
\begin{equation}
\label{FHGF_entry}
\begin{aligned}
\big[\mathbf F^{\mathcal H}\boldsymbol{\Gamma}^{\kappa_p}\mathbf F\big]_{n,m}
& =\frac{1}{N}\sum_{k=0}^{N-1}
e^{-j \tfrac{2\pi\kappa_p}{N} k}\,e^{j \tfrac{2\pi}{N} k (n-m)} \\
 & =   D_N\!\Big(\frac{n-m-\kappa_p}{N}\Big),\\
\end{aligned}
\end{equation}
where $D_N(x)$  denotes the length-$N$ Dirichlet kernel with its expression given by 
\begin{equation}
\label{eq:Dirichlet}
D_N(x)\triangleq \frac{1}{N}\sum_{k=0}^{N-1}e^{j2\pi k x}
= e^{j\pi (N-1)x}\frac{\sin(\pi N x)}{N\sin(\pi x)}.
\end{equation}
Based on (\ref{FHGF_entry}), one has
$ \mathbf F^{\mathcal H}\boldsymbol{\Gamma}^{\kappa_p}\mathbf F =\sum_{k=0}^{N-1} D_{N} \Big ( \frac{k-\kappa_p^{\text{F}} }{N} \Big) \boldsymbol \Pi^{k+\kappa_p^{\text{I}}}$. 

Upon substituting (\ref{Hpt}) and \eqref{FHGF_entry}   into (\ref{heff}), one has the input-output relationship under FDFD.  The detailed expression of $\mathbf H_{p}$   is given by (\ref{finalH})  at the top of  {the next page}.  
In the following, we develop a detailed analysis of \eqref{finalH} by considering three cases.

\textit{Case-I)  Integer Delay \& Fractional Doppler (IDFD):} One can verify that (\ref{finalH}) reduces to    (30) of  \cite{bemani2023affine}, which has also been widely investigated in existing  treatises \cite{bemani2023affine,rou2024orthogonal,
 luo2026chirp}.

\textit{Case-II)    Fractional Delay \& Integer Doppler (FDID): }  After digital sampling,   the $p$th path's fractional delay causes  $N$  {delay taps} in the digital domain with the gain of  the $k$th tap given by $\mathcal F_p^k(v)$.  {It is well known} that \cite{bemani2023affine}
\begin{equation}
\label{DnUBound}
   \vert  D_N(x) \vert = \left \vert \frac{\sin(\pi N x)}{N\sin(\pi x)} \right   \vert \leq \frac{N-1}{N}  \vert \cos(\pi x ) \vert + \frac{1}{N}.  
\end{equation}
  Note that $ \cos(\pi x) $ attains its maximum value at $x = 0$, decreases monotonically over the interval $[0, \tfrac{1}{2})$, and then increases over $[ \tfrac{1}{2}, 1)$. Therefore, $\mathcal{F}_p^k(v)$ achieves its peak value at $k = 0$ and gradually decreases, as $k$ moves away from $0$ toward $N/2$, following a periodic and symmetric trend. Fig.~ \ref{FracDel}(a) and  (b) illustrate  {examples} of $|\mathcal{F}_p^k(v)|$ for $\kappa_p^{\text{F}} = 0.1$ and $\kappa_p^{\text{F}} = 0.5$, respectively, where   $N = 32$ is assumed. As observed from Fig.~ \ref{FracDel}, the fractional delay results in energy leakage across the $N$ subcarriers. In particular, the case of $\kappa_p^{\text{F}} = 0.5$ represents the worst scenario, where the leakage effect is most severe.  The corresponding $\mathbf{H}_p$ with  $\kappa_p^{\text{F}} = 0.1$ and $\beta_p  = 0$ is also shown in Fig. \ref{ChR}(b). For $\beta_p = 0$, the fractional delay taps are well separated in the effective channel matrix. 
Note that $\mathbf{H}_p[u,v]$ has the same magnitude as that of $\mathbf{H}_p[<u+1>_N,<v+1>_N]$, but   differs in phase, which is caused by the term $e^{j \frac{2 \pi}{N} (N c_1 k^2 - v k)}$. 


\textit{Case-III)  FDFD:} In (\ref{finalH}), $\mathcal D_p^k(u,v)$ models  the integer  delay and the Doppler shift. Based on    (\ref{DnUBound}), $\vert \mathcal D_p^k(u,v) \vert $ achieves its peak value,  when $v = \langle u+ \text{loc}_p^k\rangle _{N}$, where $\text{loc}_p^k=2Nc_1(k+\kappa_p^{\text{I}})+ \beta_p^{\text{I}}$.   {For FDFD channels, the delay and Doppler components} become mutually coupled, resulting in significant overlap within the channel matrix. 
  Fig.~ \ref{ChR}(c) and Fig. \ref{ChR}(d) illustrate  the effective channel matrix $\mathbf{H}_p$ for $\kappa = 0.1, \beta = 0.2$ and $ \kappa = 0.5, \beta = 0.2$, respectively. Note that Fig.~\ref{ChR} illustrates a single-path channel matrix. In practical scenarios associated with $P > 1$, where each path may exhibit   different delay and Doppler shifts, the effective channel matrices    overlap substantially, posing serious challenges for accurate channel estimation in the DAFT domain.

\vspace{-1em}
\subsection{The Proposed   CDS-CE}
\label{PSCE}

Due to the presence of FDFD, direct channel estimation in the DAFT domain is challenging, since the multipath channels are no longer separable. For channel estimation, the goal is to recover the time shift, frequency shift, and channel coefficient of each path, i.e., $ \tau_p, \nu_p^l $ and $ h_p $, $\forall p, \forall l$. From \eqref{Mix1}, it is possible to estimate $ \tau_p, \nu_p^l $ and $ h_p $ in the TD based on the dechirp technique using coherent estimation. Unfortunately, for large values of the center frequency, we have $\left  [\left(\frac{1}{2}+q\right)B - f_c^l \right ] \tau_p \gg 1$, which introduces phase ambiguity in the estimation of $ {\theta}_{p,q}^{l} $. Therefore, in the proposed CDS-CE scheme, the estimation problem is divided into two stages. In the first stage, $ \tau_p $ and $ \nu_p^l $ are estimated via  {noncoherent} estimation using frequency recovery algorithms that utilize TD dechirping sensing techniques. Then, the channel coefficients are further recovered in the DAFT domain. In the following, we detail the proposed CDS-CE scheme.

As discussed in the previous section, FDFD causes energy leakage that decays from the peak value. Hence, truncation is applied to both the delay and Doppler leakage components. Let $\xi_{\nu}$ and $\xi_{\tau}$ denote the truncation lengths associated with the Doppler and delay components, respectively. Then, the input–output relationship of \eqref{heff} is rewritten as 
\begin{equation}
\label{IOt}
          y[u] =  
         \sum_{p= 0}^{P_c-1}    h_p  \mathbf v_p^{(\kappa_p, \beta_p)}[u]   +  \tilde {\mathbf {w}}[u],  
\end{equation}
where we have
\begin{equation}
\label{vpu}
\begin{aligned}
        & \mathbf v_p^{(\kappa_p, \beta_p)}[u]= \\
        & \sum_{k= <-\xi_{\tau}>_{N}}^{ \xi_{\tau} }    \Bigg \{ D_{N} \Big ( \frac{k-\kappa_p^{\text{F}} }{N}  \Big)    \sum_{v= (u+\text{loc}_p^k- \xi_{\nu})_{N} }^{v= (u+\text{loc}_p^k+ \xi_{\nu})_{N} }   e^{j2\pi c_2(v^2-u^2) }  \\
          & e^{ j \frac{2 \pi}{N}  \left (Nc_1 (k+\kappa_p^{\text{I}})^2-  v(k+\kappa_p^{\text{I}}) \right )  }     D_{N}\Big (    \frac{v-\! u-\! \text{loc}_p^k- \beta_p^{\text{F}}}{N}  \Big) x[v]  \Bigg \}.
\end{aligned}
\end{equation}

\begin{table}  
\footnotesize
     \caption{Simulation Parameters}
    \centering
    \begin{tabular}{c|c}
    \hline
     \hline
       \textbf{ System Parameters}  &  \textbf{ Values}  \\
        \hline
        \hline
        Center frequency & $3.3$ GHz \\
            \hline
        CA bands &   \makecell{ n78 (3300 MHz – 3800 MHz) \\ n79 (4400 MHz – 5000 MHz)} \\
        \hline
        Bandwidth & 10 MHz. \\
        \hline
            Q   &  $2$ \\
          \hline
                 CPP length   &  $24$ \\
                 \hline
                 Subcarrier spacing  &  $ \Delta f = 15$  kHz\\
                   \hline
                  Frame length  &  \makecell{$128$  symbols  for sensing, \\ and  $14$  symbols  for CDS-CE.}\\
   \hline
      SPS-to-data power ratio   &  $20$ dB \\
        \hline
        GI length   &  $N_{\text{GB}}=40$ \\
   \hline
       Target range \& speed & \makecell{$R_{p} \in \mathcal U [200,  300]$ m, \\ $V_{p} \in \mathcal U [120,  300]$ km/h }    \\
   \hline
    ADC sampling rate &   $  2\left(2\alpha R_{\max}/c+f_{\text d,\max}\right)    $  \\
 \hline
             Communication  user speed   &   $v_{\text{ue}} \in \{160,600\}$ km/h    \\
 \hline
 Doppler model &   Jakes' model \\
  \hline
 FDFD truncation length & $\xi_{\tau}=15$, $\xi_{\nu}=2$\\
 \hline
 \hline
    \end{tabular}
    \label{simparaa}
 \end{table}

Based on \eqref{IOt}, the number of guard subcarriers for communication should be  {designed to ensure} that the interference between data and pilot  is negligible, i.e.,
\begin{equation}
    N_{\text{GB}} \geq 2Nc_1 \left (\xi_{\tau}+\kappa_{p_{\max}}^{\text{I}} \right)+\xi_{\nu}.
\end{equation}
Then, the input-output relationship based on the pilot signal \eqref{IOt}    can be rewritten in   matrix form as 
\begin{equation}
\label{pilotv}
    \mathbf y_{\text{p}} =  [\mathbf v_0, \mathbf v_1,\ldots, \mathbf v_{P-1}] \mathbf h+ \tilde {\mathbf {n}}=  \mathbf V \mathbf h+ \tilde {\mathbf {n}}_{\text{p}},
\end{equation}
where $\mathbf v_p$ is the pilot and guard subcarrier section  associated with    $\mathbf v_p^{(\kappa_p, \beta_p)}$ defined in (\ref{vpu}), and $\mathbf h = [h_0,h_1,\ldots, h_{P_c-1}]^{\mathcal T}$.  Given (\ref{pilotv}), the channel  {coefficients} can be estimated in the DAFT domain as 
\begin{equation}
\label{EstH}
    \hat{\mathbf{h}}_{\text{MMSE}} 
    = \mathbf{R}_h \mathbf{V}^H
    \left( \mathbf{V}\mathbf{R}_h \mathbf{V}^H + \sigma_n^2 \mathbf{I}_{N_p} \right)^{-1}
    \mathbf{y}_{\text{p}},
\end{equation}
where $\mathbf{R}_h = \mathbb E[ \mathbf{h}\mathbf{h}^{\mathcal H}]$ denotes the covariance matrix of $\mathbf{h}$.  Finally, the estimated effective channel can  {be reconstructed} by $\hat {\mathbf H}_{\text{eff}} =  \sum_{p=0}^{P_c-1} \hat{{h}}_p{\mathbf H}_{p}$  and ${\mathbf H}_{p}$  is defined in   \eqref{finalH}.


\vspace{-0.8em}
\section{Simulation Results}

This section evaluates the sensing and communication performance of the proposed AFDM-ISAC with CA. Unless  otherwise stated, the
detailed simulation parameters  {in} Table  \ref{simparaa} are employed.  For the proposed AFDM-ISAC scheme,   a dedicated chirp subcarrier  {and} $2N_{\text{GB}}$ guard  {subcarriers} are employed for sensing and channel estimation.     {For a fair comparison, similar to the CA-assisted OFDM-ISAC scheme in \cite{WeiCAISAC}, we consider a staggered pilot-based ISAC signal. Specifically, the OFDM-ISAC benchmark employs $N_{\text p}=2N_{\text{GB}}+1$ sensing pilot subcarriers, i.e., exactly the same number of resource elements as the SPS-plus-guard structure of the proposed scheme. These pilots are uniformly distributed across the  bandwidth occupied. OFDM sensing is performed by dividing the received pilot symbols by the corresponding transmitted pilots in the frequency domain, followed by the same 2D-FFT/CFAR processing \cite{WeiCAISAC,ye2025reconstructed}. For communication performance evaluation, we adopt the approximate maximum likelihood (AML) method proposed in \cite{bemani2023affine} as a benchmark. The exhaustive search step size for both fractional delay and fractional Doppler is set to $0.02$. In addition, the perfect channel state information (CSI) case is included as a performance upper bound.}

\vspace{-0.8em}
\subsection{Complexity Analysis}
{We first compare the implementation and computational complexity of the proposed ISAC scheme to those of  conventional OFDM-ISAC schemes. Conventional OFDM-ISAC receivers generally require a wideband ADC operating at approximately the full system bandwidth, i.e., $f_{\mathrm{ADC}}\approx B$. By  contrast, owing to the analog-domain dechirping operation, the proposed receiver only has to sample the low-frequency beat signal. Hence, the   ADC sampling rate required is substantially lower than the system bandwidth. Moreover, the proposed analog-domain sensing receiver suppresses the strong transmitter leakage through dechirping followed by LPF, thereby avoiding the costly full-duplex self-interference cancellation. Finally, the proposed CA schemes aggregate the observations before 2D-FFT processing. For Type-I CA, fusing $L$ bands and $Q$ chirp segments requires only $\mathcal O(LQMM_{\mathrm A})$ complex additions, where $M_{\mathrm A}$ denotes the number of ADC samples collected within each chirp segment. By  contrast, separate processing requires $LQ$ 2D-FFTs. Thus, the proposed aggregation significantly reduces the sensing-processing complexity.
}

{For the proposed CDS-CE scheme, the additional implementation overhead mainly arises from the dechirping-based receiver, as discussed above. The computational complexity of channel-gain estimation in \eqref{EstH} is approximately $\mathcal{O}(N_{\text{GB}}^3)$. By  contrast, the AML estimator can reuse the communication hardware, but requires an exhaustive $2P_c$-dimensional grid search with complexity scaling as $\mathcal{O}\big((N_\tau N_\nu)^{P_c}\big)$, where $N_\tau$ and $N_\nu$ denote the numbers of delay and Doppler grid points, respectively. In general,  AML becomes computationally intractable for $P_c>4$. }

\begin{figure}[tbp]
    \centering
    \includegraphics[width=0.7\linewidth]{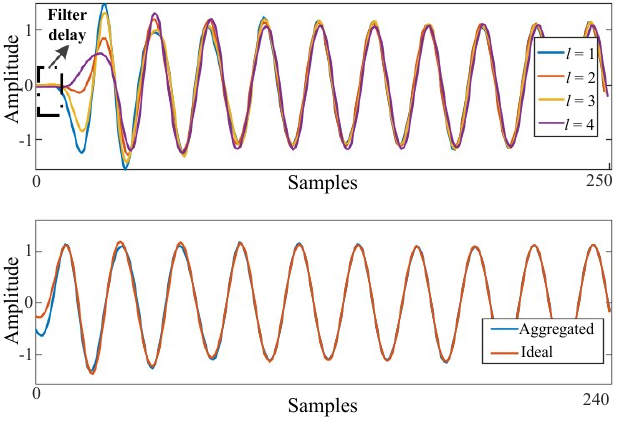}
    \caption{Illustration of the dechirped signal for different  {numbers} of CA bands.}
    \label{CADECHIRP}
   \vspace{-1.5em}
\end{figure}
\begin{figure} [tbp]
	\centering
	\begin{subfigure}{0.4 \textwidth}
\includegraphics[width=1\linewidth]{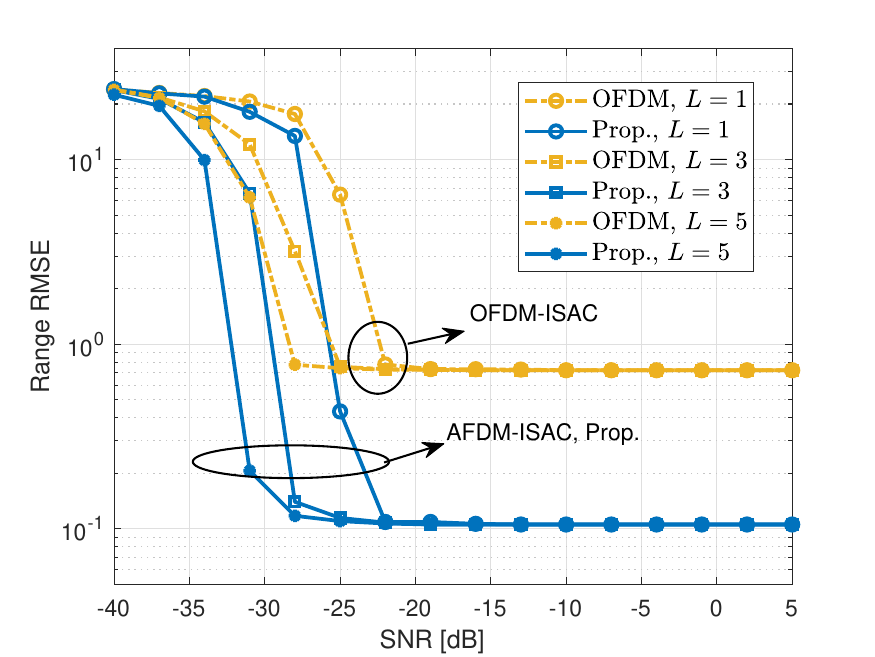}
    \caption{Range RMSE.}
 	\end{subfigure}
	\begin{subfigure}{0.4 \textwidth}
\includegraphics[width=1\linewidth]{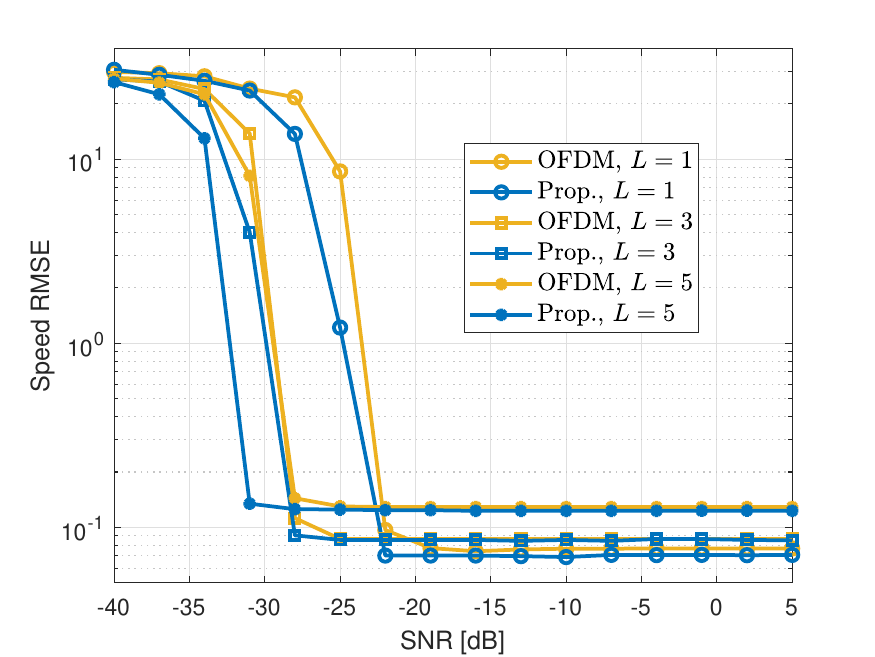}
    \caption{Speed RMSE.}
		\vspace{-0.2em}
	\end{subfigure}
                 \caption{Range- and velocity- RMSE of the proposed AFDM-ISAC  with CA (Intra-band CA).}
                     \label{IntraNonCont}
                        \vspace{-1.5em}    
\end{figure}

\vspace{-0.8em}
\subsection{Sensing Performance}

We first present the dechirped signal after the LPF for $L=4$, as shown in Fig.~\ref{CADECHIRP}. In particular, the n78 band, as shown in Table \ref{simparaa}, with contiguous CA is considered as an example. It can be observed that the dechirped signals corresponding to different CA bands exhibit slightly different initial phases. However, after applying the proposed Type-I CA aggregation scheme, the aggregated dechirped signal closely aligns with the ideal signal, indicating the efficiency of the proposed aggregation method.

Fig.~\ref{IntraNonCont} shows both the range and speed root MSE (RMSE) of the proposed AFDM-ISAC using intra-band non-contiguous CA, where the $L$ CA bands are randomly generated within n78. A 2D-FFT-based processing scheme using CFAR detection and $5$- {fold} FFT oversampling is considered. The main observations are summarized as follows:
 \begin{itemize}
     
    \item For both range and velocity RMSE, higher performance gains are observed in the intermediate RMSE region, i.e., approximately $[10^{-1}, 10^{1}]$, as $L$ increases. Beyond this region, an error floor appears. This behavior is mainly due to the limited resolution imposed by the 2D-FFT-based estimation, even when oversampling is employed.

\item The proposed AFDM-ISAC scheme achieves a significant improvement in range-RMSE compared to the OFDM-ISAC scheme. This is mainly attributed to the distortion introduced by practical pulse-shaping filters and the interference from communication subcarriers in the OFDM-based design.

\item The proposed schemes achieve similar range-RMSE for $L=1,3,5$ at high SNR. However, a higher velocity-RMSE is observed for $L=5$ compared to $L=1$ and $L=3$. Indeed, this behavior is not unexpected, as the carrier frequency mismatch becomes more pronounced upon aggregating CA bands that are farther apart, thereby degrading the velocity estimation accuracy.
 
 \end{itemize}

\begin{figure}
	\centering
	\begin{subfigure}{0.4  \textwidth}
\includegraphics[width=1\linewidth]{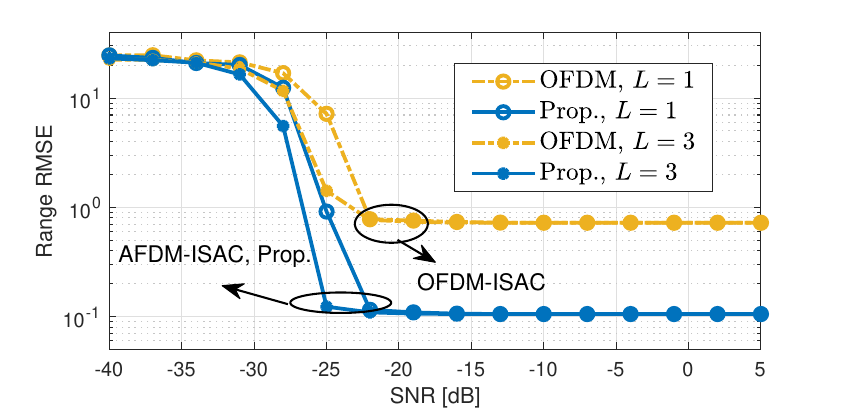}
    \caption{Range RMSE.}
				\vspace{-0.1em}
	\end{subfigure}
	\begin{subfigure}{0.4 \textwidth}
\includegraphics[width=1\linewidth]{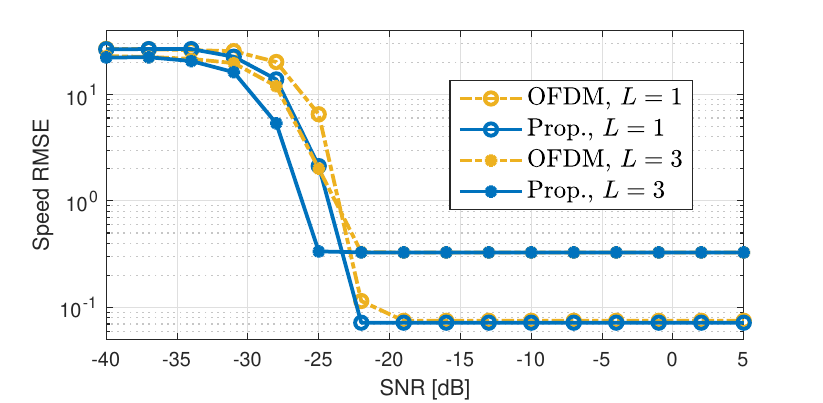}
    \caption{Speed RMSE.}
		\vspace{-0.1em}
	\end{subfigure}
                 \caption{Range- and velocity-RMSE of the proposed AFDM-ISAC  with CA (Inter-band CA).}
                        \vspace{-1.5em}
                        \label{InterCA}
\end{figure}
\begin{figure}
    \centering
    \includegraphics[width=0.9\linewidth]{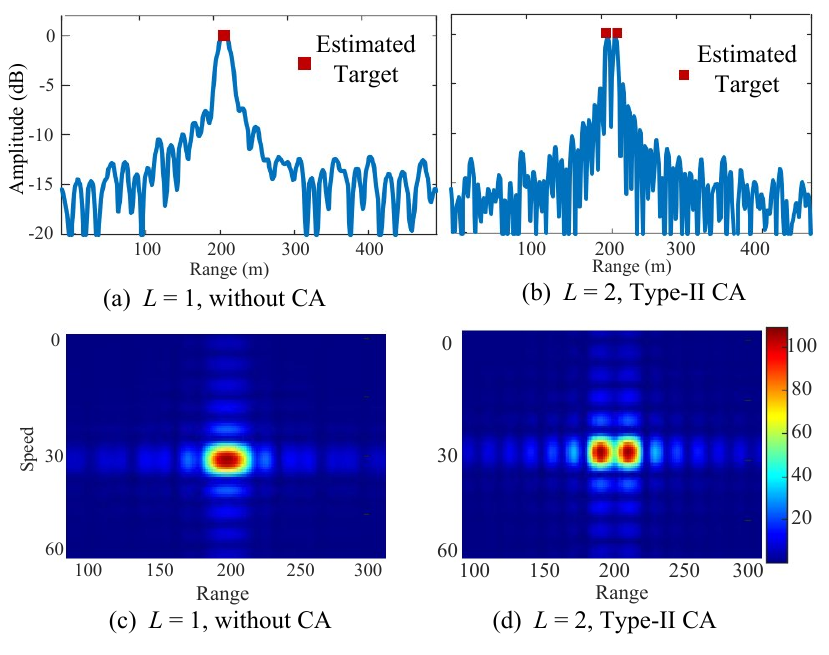}
    \caption{Illustration of the range FFT and 2D-FFT map for the proposed Type-II CA.}
    \label{rfftmap}
    \vspace{-1.5em}
\end{figure}

\begin{figure*}[!htbp]
    \centering
\includegraphics[width=0.95\linewidth]{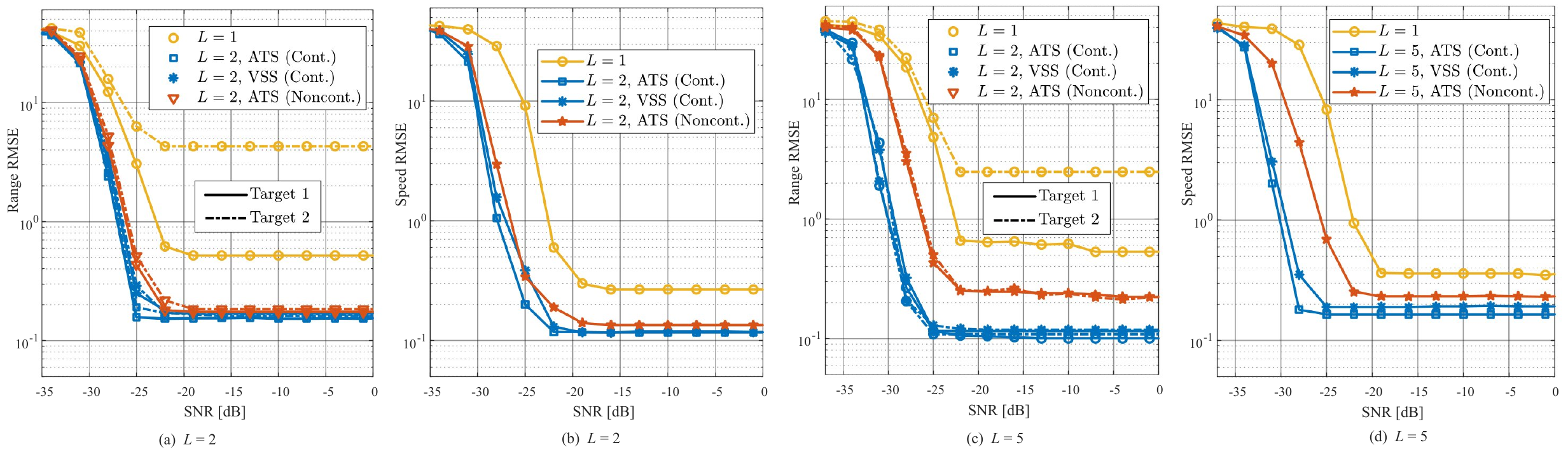}
    \caption{RMSE of estimated targets.}
    \label{RMSECA2}
       \vspace{-1.5em}
\end{figure*}

\begin{figure}[!htbp]
    \centering
\includegraphics[width=0.8\linewidth]{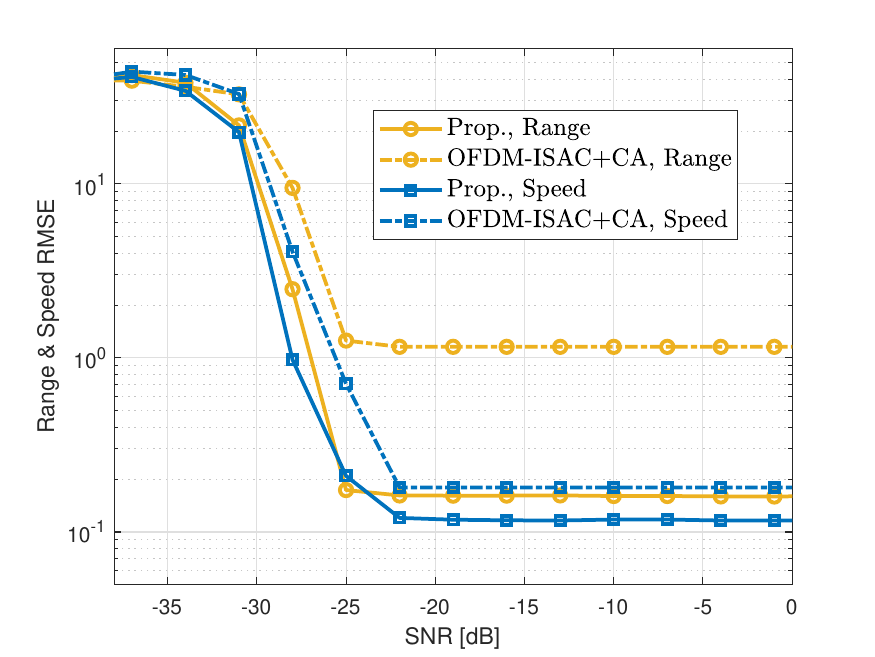}
    \caption{{Range- and velocity-RMSE of the proposed  ATS  and OFDM-ISAC with CA. }}
    \label{AFDM_OFDMCAii}
    \vspace{-1.5em}
\end{figure}

Fig.~\ref{InterCA} further compares the range- and velocity-RMSE performance under inter-band CA for $L=3$, specifically between the n78 and n79 bands. The carrier frequency $f_c^1$ is set to $f_c^1 = 3.5$ GHz, while $f_c^2$ and $f_c^3$ are randomly generated within the n79 band, i.e., $4.4\text{ GHz} \leq f_c^2, f_c^3 < 5\text{ GHz}$. As observed from the figure, when the carrier frequency mismatch becomes more pronounced, a higher error floor appears for $L=3$. Nevertheless, the proposed Type-I CA scheme still improves the range-RMSE and achieves velocity-RMSE gains in certain SNR regions.

We now evaluate the sensing performance of the proposed Type-II CA scheme. Fig.~\ref{rfftmap} illustrates the range FFT and the corresponding 2D-FFT map for a pair of targets with range and speed given by $(200 \text{ m}, 30 \text{ m/s})$ and $(208 \text{ m}, 30 \text{ m/s})$, respectively. For $L=2$, the proposed VSS scheme with contiguous CA within the n79 band is considered. As shown in Fig.~\ref{rfftmap}, the proposed VSS effectively improves the range resolution, and the   targets are clearly separated.
We next present the range and velocity RMSE performance of the proposed Type-II CA schemes, as shown in  {Fig.}~\ref{RMSECA2}. The carrier frequency of the first band is set to $f_c^1=3.3$ GHz, while the remaining bands are divided into CA bands with a bandwidth of $10$ MHz each. The curves labeled “Cont.” and “Noncont.” denote contiguous and noncontiguous CA, respectively.

In Fig.~\ref{RMSECA2}(a) and (b), the two targets share the same velocity, but are separated by $\Delta R = 7$ m. For the noncontiguous CA case, the first and third bands are selected, corresponding to a frequency separation of $f_c^2 - f_c^1 = 20$ MHz. In Fig.~\ref{RMSECA2}(c) and (d), we further consider a higher number of CA bands, i.e., $L=5$, and assume $\Delta R = 4$ m. For the noncontiguous CA scenario, the sub-bands indexed by $[1,2,4,5,7]$ are adopted.
As observed from  {Fig.}~\ref{RMSECA2}, without CA, the targets exhibit poor RMSE performance, since they cannot be confidently resolved. By  contrast, the proposed Type-II CA scheme significantly improves the range resolution, leading to enhanced RMSE performance. It should be noted that the proposed ATS slightly outperforms VSS in terms of RMSE under the noncontiguous CA scenario. This is because not all CA bands can utilize the optimal chirp subcarrier, i.e., the $N/2$th subcarrier, for sensing,  {resulting} in a certain SNR degradation compared to ATS. The performance gap becomes more pronounced as $L$ increases, as illustrated in Fig.~\ref{RMSECA2}(c) and (d) for the case of $L=5$. Nevertheless, the proposed Type-II CA schemes still significantly outperform the scenario of $L=1$.

{Finally, we compare the range- and velocity-RMSE performance of the proposed Type-II CA scheme to that of CA-assisted OFDM-ISAC. Specifically, the ATS scheme is considered, and the RMSE is averaged over the two targets. The simulation parameters are the same as those in Fig.~\ref{RMSECA2}(a) and (b). For the OFDM-ISAC baseline, the pilot observations from different CA bands at the $i$th symbol are concatenated to expand the effective sensing bandwidth. As shown in Fig.~\ref{AFDM_OFDMCAii}, the proposed Type-II CA scheme significantly outperforms CA-assisted OFDM-ISAC, particularly in terms of range RMSE. The  {ECCS} constructed by ATS effectively converts the fragmented AFDM chirp segments across different CA bands into a continuous sensing aperture, while accounting for the guard bands and CPP-induced discontinuities. Moreover, under large Doppler, the OFDM pilot observations suffer from ICI induced by the data subcarriers, which further degrades delay estimation. By  contrast, the performance gap in velocity RMSE is less pronounced, since velocity resolution is mainly determined by the width of coherent processing interval.}

\vspace{-1em}
\subsection{Communication Performance}

\begin{figure}[htbp]
    \centering
\includegraphics[width=0.82\linewidth]{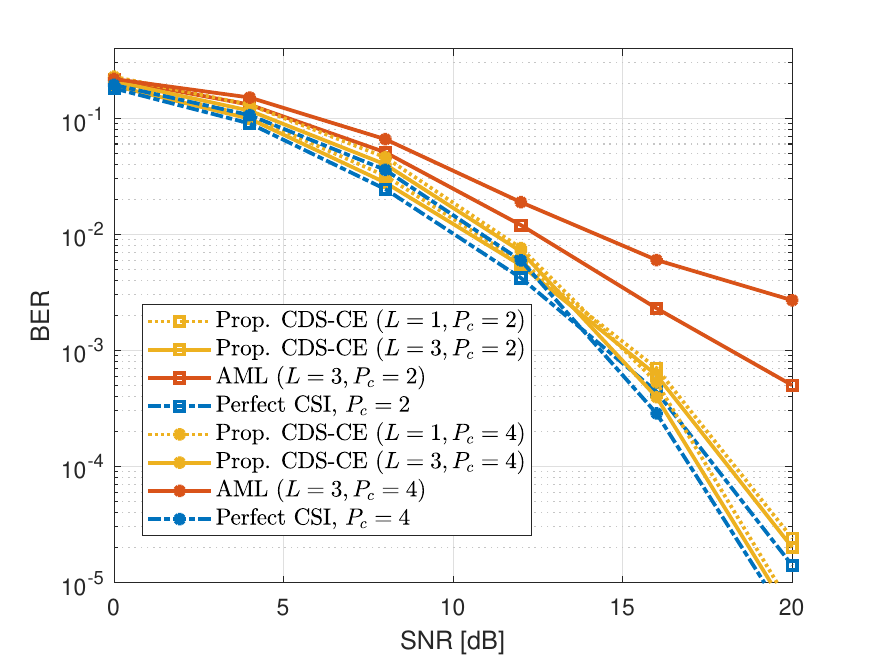}
    \caption{BER comparisons for different channel estimation schemes.}
    \label{bercdce}
    \vspace{-1.5em}
\end{figure}

\begin{figure}[htbp]
    \centering
\includegraphics[width=0.8\linewidth]{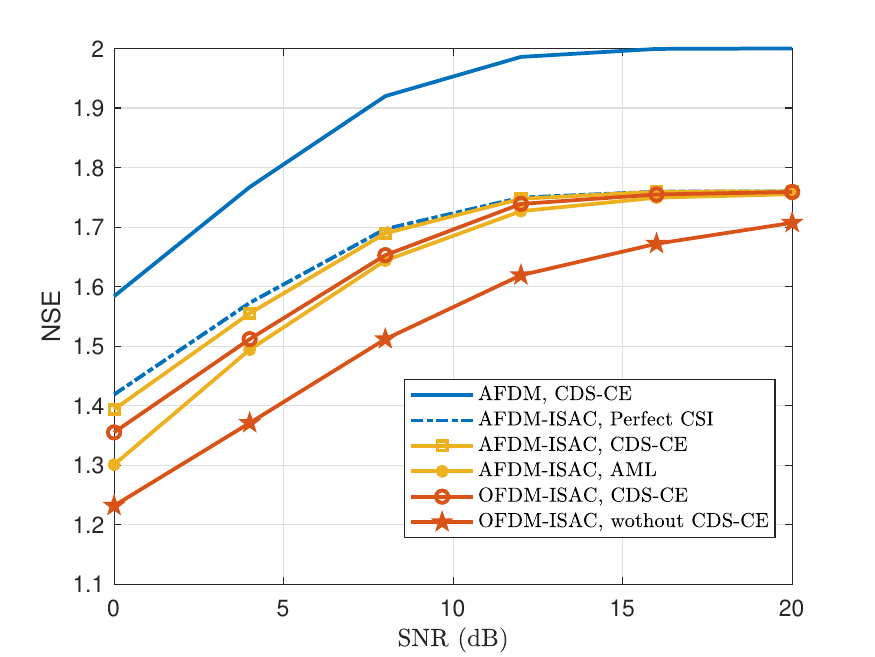}
    \caption{{NSE comparisons for AFDM and OFDM with different channel estimation schemes.}}
    \label{ncecom}
    \vspace{-1.5em}
\end{figure}

\begin{figure}[tbp]
    \centering
\includegraphics[width=0.76\linewidth]{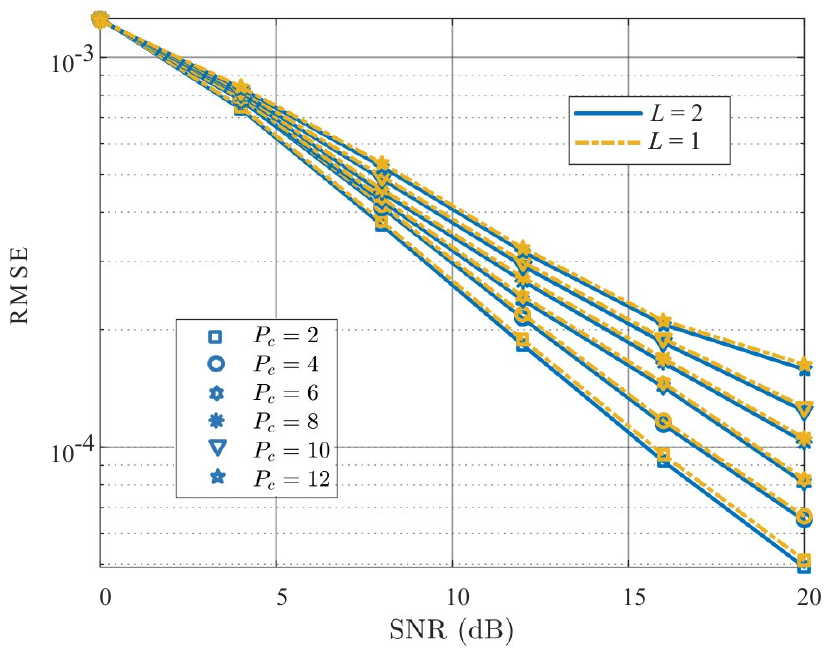}
    \caption{RMSE of the estimated channels.}
    \label{RMSEH}
\vspace{-1.5em}
\end{figure}

\begin{figure}[tbp]
    \centering
\includegraphics[width=0.86\linewidth]{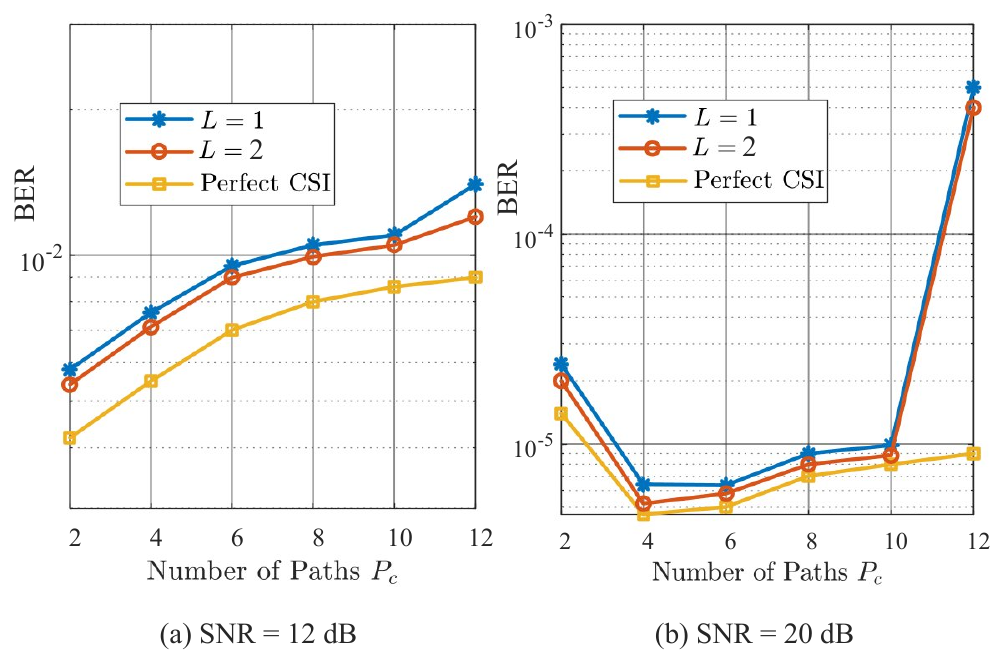}
    \caption{BER performance  {vs.} different  {numbers} of paths.}
    \label{BERpATH}
    \vspace{-1.5em}
\end{figure}

In this section, we evaluate the communication performance of the proposed CDS-CE scheme in the AFDM-ISAC system, as illustrated in Figs.~\ref{bercdce}–\ref{BERpATH}. 
We aim to evaluate the efficiency of the proposed CDS-CE scheme. Hence, the throughput gain introduced by CA is omitted, as it is straightforward and well understood. The number of channel paths is denoted by $P_c$, with each path assigned a power of $1/\sqrt{P_c}$. The normalized delay, including both integer and fractional components, is randomly generated within $[0, 15]$. The maximum fractional Doppler is set to $f_d^{\max}=0.4$, and the Jakes Doppler spectrum is adopted. Additionally, the number of symbols employed for Doppler and delay estimation is $14$.

Fig.~\ref{bercdce} shows the BER performance of the proposed CDS-CE scheme for $P_c=2$ and $P_c=4$. As observed, the proposed CDS-CE approaches the   BER performance of the perfect CSI case. No significant difference is observed between $L=1$ and $L=2$, which is reasonable, since for a single CA configuration, both the delay and Doppler parameters can be accurately estimated. Interestingly, in the presence of FDFD, the BER performance of $P_c=4$ is only better than that of $P_c=2$ in the high-SNR region due to the diversity gain. Moreover, the proposed CDS-CE significantly outperforms the AML scheme  of  \cite{bemani2023affine}.{ In particular, it provides an approximately $7$ dB gain over the AML-based channel-estimation scheme at $\mathrm{BER}=5\times10^{-3}$ for $P_c=4$.  It is worth noting that the computational complexity of the AML scheme escalates exponentially with $P_c$ and becomes intractable when $P_c>4$.}

 We then compare the    communication efficiency   of AFDM-ISAC and OFDM-ISAC systems. Specifically, we consider the normalized spectral  efficiency (NSE) of $L$ bands, defined as
\begin{equation}
    \label{NSE}
    \begin{aligned}
            \eta_{\text{NSE} }&=    \frac{ \text{Number of communication subcarriers}}{  \text{Total number of subcarriers}} (1-\text{SER})\log_2(M) \\
   &=  \frac{N- 2N_{\text{G}}  -1 }{N}(1-\text{BER})\log_2(M).    
    \end{aligned}
\end{equation}
Fig.~\ref{ncecom} shows the NSE of the AFDM- and OFDM-ISAC systems relying on $L=3$ and $P_c=4$, under the same channel settings as in Fig.~\ref{bercdce}. For OFDM-ISAC, we consider both the proposed CDS-CE and the conventional MMSE channel estimator, labeled as ``without CDS-CE''. The curve ``AFDM, CDS-CE'' denotes the communication-only AFDM case, where all subcarriers are allocated to communication. As it can be seen, AFDM associated with the proposed CDS-CE achieves an NSE close to the perfect-CSI case and significantly outperforms the OFDM counterparts. In particular, without CDS-CE, the NSE of OFDM degrades considerably due to   ICI and the resulting channel-estimation errors.  Compared with the communication-only AFDM case, an NSE loss of approximately  $3$ bits/channel use   is observed due to the allocation of subcarriers for sensing.

\color{black} 
Fig.~\ref{RMSEH} further presents the RMSE of the estimated effective channel, i.e., $ \Vert \mathbf H_{\text{eff}} - \hat{\mathbf H}{\text{eff}} \Vert / \Vert \mathbf H{\text{eff}} \Vert $. As expected, the channel estimation RMSE increases with $P_c$, since a higher number of propagation paths introduces higher estimation uncertainty. Additionally, CA has a negligible impact on the channel estimation performance, indicating that a single CA band is sufficient for accurate delay and Doppler estimation. Fig.~\ref{BERpATH} illustrates the BER performance of the proposed CDS-CE scheme versus the number of propagation paths. The proposed CDS-CE maintains reliable channel estimation performance even when the number of paths is relatively high, i.e., $P_c \leq 10$. However, when $P_c \geq 12$, the performance deteriorates due to degraded sensing-assisted delay and Doppler estimation, which consequently leads to inaccurate channel recovery.

\vspace{-1em}
\section{Conclusions}
This paper proposed a CA-assisted AFDM-ISAC scheme
that exploits the inherent chirp structure of AFDM through
FMCW-inspired dechirping.  An FMCW-inspired dechirping receiver was first developed for efficient delay--Doppler estimation. To enhance sensing over fragmented spectrum, a pair of CA schemes  {was} proposed: Type-I improves sensing accuracy through multi-band echo aggregation, while Type-II enhances range resolution by constructing an  {effective continuous} chirp signal set to enlarge the sensing bandwidth. In addition, a cross-domain sensing-assisted channel estimation scheme was developed for fractional delay and Doppler channels, combining TD delay--Doppler estimation with DAFT-domain channel-gain recovery. Simulation results demonstrated significant improvements in sensing accuracy, range resolution, and channel estimation performance with reduced hardware overhead, highlighting the potential of CA-enabled AFDM-ISAC for future high-mobility communications over fragmented spectrum.

{Future work includes peak-to-average power ratio reduction for AFDM-ISAC, operation in higher-frequency bands, and MIMO-ISAC extensions for angle estimation. In addition, practical aspects, such as software-defined radio-based prototyping and experimental validation, as well as the characterization of hardware impairments   including TO, CFO, phase noise, and PA nonlinearity,  can also be investigated \cite{lu2024integrated,10726912}. }

\vspace{-0.8em}
\bibliographystyle{IEEEtran}
\bibliography{ref}

\end{document}